\documentclass{aa}
\usepackage{graphicx}
\usepackage{amsmath}
\usepackage{amssymb}
\usepackage{natbib}
\bibpunct{(}{)}{;}{a}{}{,}
\newcommand{\Ds}{D_{\mathrm{s}}}
\newcommand{\Dd}{D_{\mathrm{d}}}
\newcommand{\Dds}{D_{\mathrm{ds}}}
\newcommand{\map}{M_{\mathrm{ap}}}
\newcommand{\parcm}{\;{\mathrm{arcmin}}^{-2}}
\newcommand{\e}{\varepsilon}
\newcommand{\nup}{$\nu_{\mathrm{p}}\,$}
\newcommand{\nus}{$\nu_{\mathrm{s}}\,$}

\newcommand{\myarcsec}{\hbox{$.\!\!^{\prime\prime}$}}
\newcommand{\myarcmin}{\hbox{$.\!\!^{\prime}$}}
\newcommand{\myarcsecnodot}{\hbox{$\;\!\!^{\prime\prime}\;$}}
\newcommand{\myarcminnodot}{\hbox{$\;\!\!^{\prime}\;$}}
\newcommand{\fat}[1]{\mbox{\boldmath $ #1 $}}
\newcommand{\td}{\mathrm{d}}
\renewcommand{\l}{\left}
\renewcommand{\r}{\right}

\newcommand{\figref}[1]{Fig. \ref{#1}}

\begin{document}

\bibliographystyle{aa}

   \title{GaBoDS: The Garching-Bonn Deep Survey}

   \subtitle{IX. A sample of 158 shear-selected mass concentration candidates
     \thanks{Based on observations made with ESO Telescopes at the La Silla 
       Observatory}}
   \author{Mischa Schirmer\inst{1}, Thomas Erben\inst{2}, 
     Marco Hetterscheidt\inst{2}, Peter Schneider\inst{2}
   }

   \offprints{M. Schirmer}

   \institute{Isaac Newton Group of Telescopes, Calle Alvarez Abreu 70,
              38700 Santa Cruz de La Palma, Spain; 
	      \email{mischa@ing.iac.es}
	 \and Argelander-Institut f\"ur Astronomie (AIfA), 
              Universit\"at Bonn, Auf dem H\"ugel 71, 
              53121 Bonn, Germany;
   }

   \date{Received; accepted}

   \abstract{}
	    {The aim of the present work is the construction of a
	      mass-selected galaxy cluster sample based on weak gravitational 
	      lensing methods. This sample will be subject to spectroscopic 
	      follow-up observations.}
	    {We apply the mass aperture statistics and a derivative of it to 
	      19 square degrees of 
	      high quality, single colour wide field imaging data obtained 
	      with the WFI@MPG/ESO 2.2m telescope. 
	      For the statistics a family of filter functions is used
	      that approximates the expected tangential radial shear profile 
	      and thus allows for the efficient detection of mass
	      concentrations.} 
	    {We identify 158 possible mass
	      concentrations. This is the first time that such a 
	      large and blindly selected sample is published. 72 of the 
	      detections are associated with concentrations of bright 
	      galaxies. For about 22 of those we found spectra in the 
	      literature, indicating or proving that the galaxies seen are
	      indeed spatially concentrated. 15 of those were previously
	      known to be clusters or have meanwhile been secured as such. 
	      We currently follow-up a larger number of them spectroscopically 
	      to obtain deeper insight into their physical properties. The 
	      remaining 55\% of the possible mass concentrations found are not 
	      associated with any optical light, or could not be classified 
	      unambiguously. We show that those ``dark'' detections are 
	      to a significant degree due to noise, and appear
	      preferentially in shallow data.}
	    {}

\titlerunning{A sample of 158 shear-selected mass concentrations}
\authorrunning{M. Schirmer et al.}

   \maketitle

\section{Introduction}
After the recent determination of the fundamental cosmological parameters 
\citep[][]{sbd06}
a profound understanding of the dark and luminous matter distribution in the 
Universe is one of the key problems in modern cosmology. Galaxy clusters are 
at the centre of attention in this context since they indicate the largest 
density peaks of the cosmic matter distribution. Their masses can be predicted
by theory, and gravitation dominates their evolution. Thus clusters are prime
targets for the comparison of observation against theory. For this purpose
they should preferentially be selected by their mass instead of their 
luminosity in order to avoid a selection bias against underluminous members. 
Weak gravitational lensing methods select cluster candidates based solely on 
their mass properties, but this method has not yet been used systematically on 
a very large scale (some 100 or 1000 square degrees) due to a lack of  
suitable data. Only a few dozen shear-selected peaks were reported so far
\citep[see][for some
  examples]{ewm00,umf00,mhs02,dpl03,wtm01,wmt03,ses03,ses04,hett05,vle05,
  wdh05}, a very small number compared to the more than $10000$ clusters known
to date \citep[see][for 9900 cluster candidates on the northern sky within 
$z=0.1\dots 0.5$]{lcg04}.

The selection of mass concentrations using the shear caused by their weak 
gravitational lensing effect suffers from a number of disadvantages. The most
important one is the high amount of noise contributed by the intrinsic 
ellipticities of lensed galaxies. This largely blurs the view of the cosmic
density distribution, letting peaks disappear, and fakes peaks where there
actually is no overdensity. It can only be beaten down to some degree by
deep observations in 
good seeing. The other disadvantage is that any mass along the line of sight
will contribute to the signal, giving rise to false peaks. Such projection
effects or cosmic shear can only be eliminated or at least recognised if 
redshift information of either the lensed galaxies, or the matter distribution 
in the field is available. Cosmic shear can act as a source of noise 
\citep[see for example][]{mmb05}, but it has recently been shown by
\citet{msm06} that this can partly be filtered out.

\begin{figure*}[t]
\includegraphics[width=1.0\hsize]{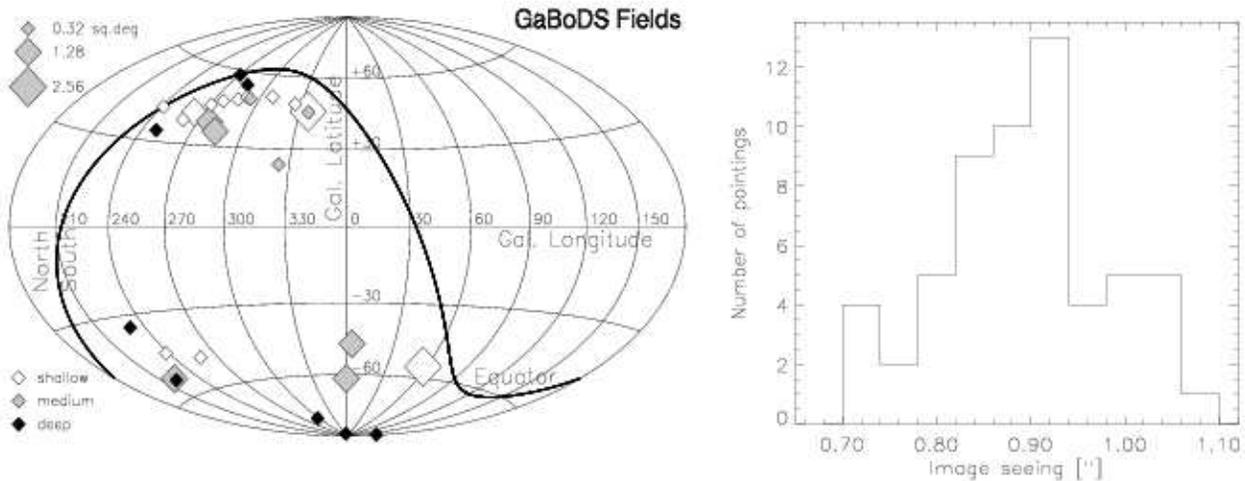}
\caption{\label{allskyseeing}{Left: Sky distribution of the GaBoDS fields. The 
    size of the symbols indicates the covered sky area (not to scale), with
    one WFI shot covering 0.32 square degrees. All fields are at high galactic
    latitude. Right: Image seeing of the 58 coadded WFI@2.2 mosaics used for
    the lensing analysis. The average seeing is 0\myarcsec86.}}
\end{figure*}

In the present work we use the aperture mass statistics ($\map$) 
\citep[][hereafter S96]{sch96_2} and a derivative of it for the 
shear-selection of density peaks, based on 19 square degrees of sky coverage.
The purpose of the work is to establish a suitable filter function for 
$\map$, and then apply it to an (inhomogeneous) set of data. The sample
returned is currently the largest sample of shear-selected cluster candidates, 
yet is dwarfed by the total number of galaxy clusters known.

The outline of this paper is as follows. In Sect. 2 we give an overview of 
the data used, concentrating on its quality and showing the usefulness for
this analysis. Section 3 contains a discussion of our implemented version of 
$\map$, particularily with regard to the chosen spatial filter functions. We 
also introduce a new statistics, deduced from $\map$. In Sect. 4 we present
and discuss our detections, and we conclude in Sect. 5.

Throughout the rest of this paper we use common weak lensing notations, and
refer the reader to \citet{bas01} and \citet{skw06} for more details and
technical coverage.

\section{Observations, data reduction, data quality}
\subsection{\label{surveysources}The survey data}
The GaBoDS was conducted with the wide field imager (hereafter WFI@2.2) of the 
2.2m MPG/ESO telescope in the $R$-band. It is to about 80\% a virtual survey,
which means most of the data was not taken by us but collected from the ESO
archive. There are five main data sources:
\begin{itemize}
\item{The ESO Distant Cluster Survey, consisting of 12 pointings that have
  been selected because $z=0.5 \dots 0.8$ cluster candidates have been 
  identified in them using colour techniques. These exposures are rather
  shallow and we are thus insensitive to such distant objects (Fig. 
  \ref{sntanh}). Therefore, these fields do not impose a selection effect and
  qualify for our survey.}
\item{Our ASTROVIRTEL\footnote{http://www.stecf.org/astrovirtel/} project for
  data mining the ESO archive, which yielded 30 random pointings. As compared 
  to the rest of the fields, many of them have unsuitable PSF properties,
  e.g. due to less careful focusing, so that we selected the 19 best
  fields. Details of this approach and its implementation are given in
  \citet{ses03} and \citet{mpb04}}.
\item{The COMBO-17 data makes for the deepest survey part. Four of the five 
  fields (S11, A901, SGP and FDF) contain known clusters.}
\item{The EIS Deep Public Survey, contributing 9 empty fields.}
\item{Our own observations of 13 empty fields.}
\end{itemize}

The final sky coverage is 18.6 square degrees, spread over 58 fields which are 
suitable for our weak gravitational lensing analysis. The overall sky 
distribution of the fields is shown in \figref{allskyseeing}, and further 
characteristics are summarised in Table \ref{gabodsfields}.

\subsection{Data reduction}
For the reduction of this specific data set we developed a stand-alone, fully
automatised pipeline (\textit{THELI}), which we made freely 
available
%\footnote{http://www.astro.uni-bonn.de/$\sim$mischa/theli}
to the community. It is capable of reducing almost any kind of optical,
near-IR and mid-IR imaging data. A detailed description of this 
software package is found in Erben et al. (2005), in which we investigate 
its performance on optical data. One of the main assets of this package is
a very accurate astrometric solution that does not introduce any 
artificial PSF distortions into the coadded images. 

The only difference between the current version of \textit{THELI} and the one
we used for the reduction of the survey data over the last years, is that for
the image coaddition \textit{EISdrizzle} was used, which is meanwhile replaced
by \textit{Swarp}. The latter method leads to a 4\% smaller PSF in the final 
image in case of superb intrinsic image seeing as in our survey. The PSF
anisotropy patterns themselves are indistinguishable between the two
coaddition methods. Since the natural seeing variations 
(\figref{allskyseeing}) in our images are much larger than those 4\%, our
analysis remains unaffected.

\subsection{Image seeing and PSF properties}
The image seeing and the PSF anisotropy are critical for weak lensing
measurements. They dilute and distort the signal and determine how well the 
shape of the lensed galaxies can be recovered. As can be seen from 
\figref{allskyseeing}, 80\% of our coadded images have sub-arcsecond 
image seeing, and 20\% are around 1\myarcsec0, with an average of 
0\myarcsec86. Thus, we reach the sub-arcsecond seeing regime very well which
is commonly regarded as mandatory for weak lensing purposes.

\begin{figure}[th]
\includegraphics[width=1.0\hsize]{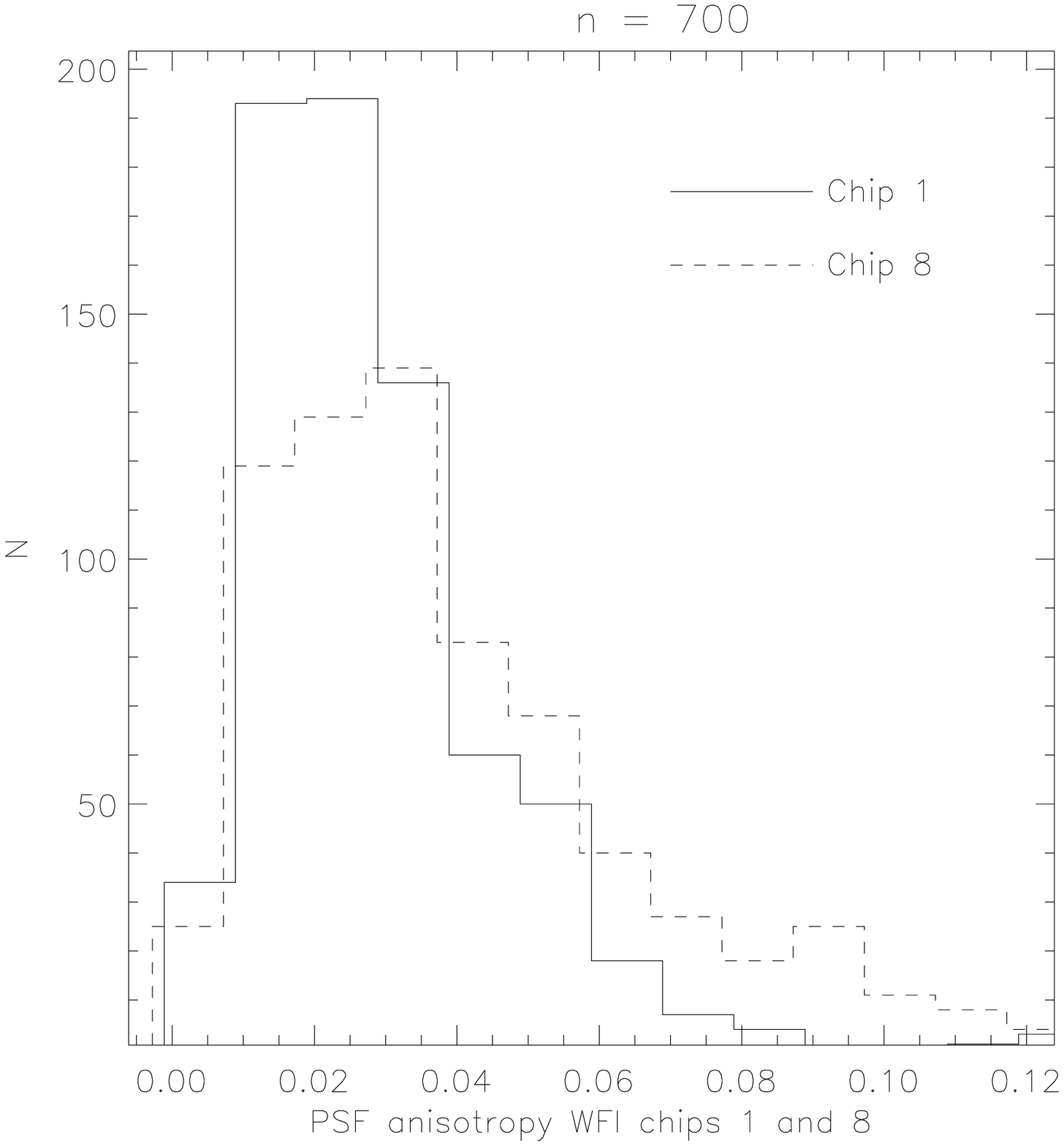}
\caption{\label{wfitoppsf}{Characteristic PSF anisotropies $\l<|\e^{*}|\r>$ 
  (modulus of the ellipticity of stars, averaged over the entire CCD) of 700
  randomly chosen exposures for CCDs 1 and 8
  of the WFI@2.2 detector mosaic. The remaining six CCDs have somewhat better
  properties. The exposure times were in the range of 300-500 seconds, and the 
  zenith distances were smaller than 40 degrees. The differences between the
  distributions arise from slightly different spatial orientations with
  respect to the focal plane (the CCDs are confined within $\pm20$ microns),
  thus responding differently to the focus. Anisotropies larger than $1-2\%$
  are mostly due to an increasingly defocused instrument. The exposures used
  for this statistics were taken by a dozen different observers spread over
  more than a year, and give an idea of the quality of the archival data.}}
\end{figure}

PSF anisotropies are rather small and usually well-behaved with WFI@2.2, which
we have demonstrated several times \citep[][for example]{ses03,ses04,esd05}. 
With a well-focussed telescope, 1\% of anisotropy in sub-arcsecond seeing 
conditions can be achieved, with a long-term statistical mode of around 2\% 
(see \figref{wfitoppsf}). Discontinuities in the PSF are largely absent across
chip borders. Slightly defocused exposures exhibit anisotropies of $3-5\%$. We
rejected individual exposures from the coaddition if one or more CCDs had an
anisotropy of larger than 6\% in either of the ellipticity components $\e_1$ 
or $\e_2$. These anisotropies arise from astigmatism, and they flip by 90
degrees if one passes from an intrafocal to an extrafocal exposure
\citep[see][for an example of this, and Fig. \ref{mosaniso_2} for a typical PSF 
anisotropy pattern in our data]{ses03}.

\begin{figure*}[t]
\includegraphics[width=1.0\hsize]{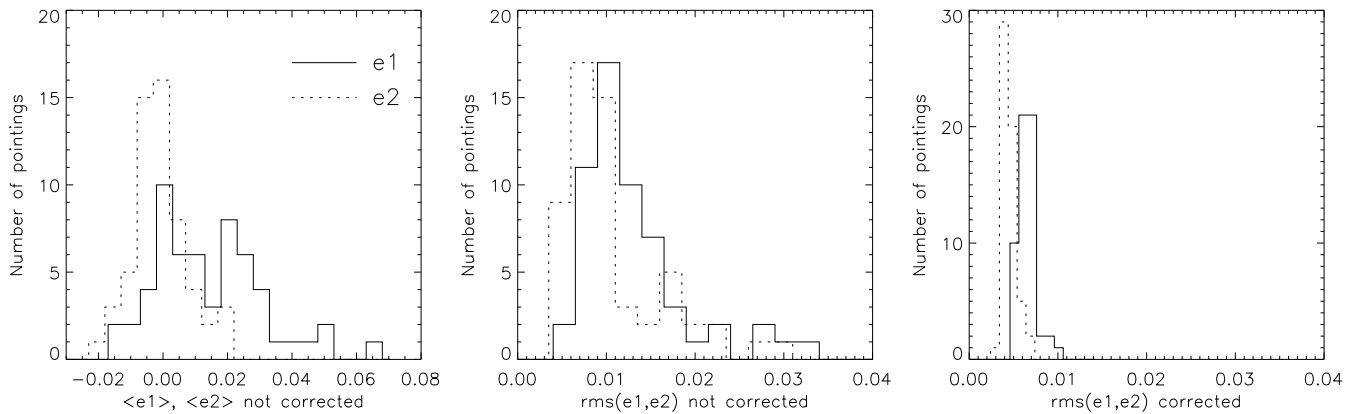}
\caption{\label{gabods_aniso}{Left: The average $\e_{1,2}$ ellipticity
    components for the PSF of each mosaic, before PSF correction. The $\e_2$ 
    component is symmetric around zero, whereas the $\e_1$ component 
    scatters broader around a value of 1.4\%. We do not show the corresponding 
    plot after PSF correction, since all $|\e_{1,2}| < 0.001$. Middle:
    The scatter of the two components around their average values. This
    indicates the deviation from a constant PSF anisotropy across the
    field and the noise level. Right: Same as in the middle, but after
    applying our PSF correction scheme. As a result, not only does 
    $\langle\e_{1,2}\rangle$ get largely removed, also their rms is 
    significantly lowered.}}
\end{figure*}

Since such intra- and extrafocal exposures are roughly equally numbered for a
larger set of exposures, anisotropies due to defocusing average out in the 
coadded images. Those have on average an anisotropy of $1-2\%$ with a similar 
amount in the rms (variation of the PSF across the field). This is illustrated 
in \figref{gabods_aniso}, where we show the combined PSF anisotropy properties
for all coadded mosaic images. The left panel shows the uncorrected mean 
ellipticity components, having $\langle\e_1\rangle\,=0.0138$ and
$\langle\e_2\rangle\,=-0.0002$. These anisotropies are small and become 
$|\langle\e_{1,2}\rangle|\,< 0.001$ after PSF correction for all mosaics.

The middle plot of \figref{gabods_aniso} shows the uncorrected rms values for 
$\langle\e_{1,2}\rangle$, i.e. the deviations of the PSF from a constant
anisotropy across the field. The rms of both components peaks around 1\% and
is reduced by a factor of 2 after PSF correction (right panel). The tail of
the distribution seen in the middle plot essentially vanishes.

To evaluate the remaining residuals from PSF correction more quantitatively, 
we calculated the correlation function between stellar ellipticity and shear 
before and after PSF correction (Fig. \ref{estar_gamma}), separately
for the various survey data sources and over all galaxy positions,
\begin{equation}
\langle \e_{1,2}^{*,{\rm pol}}\, \gamma_{1,2} \rangle = 
  1/N\; \sum_{i=1}^N \e_{i\,1,2}^{*,{\rm
  pol}}(\fat{\theta})\,\gamma_{i\,1,2}(\fat{\theta}) \;.
\end{equation}
Here $\gamma_{1,2}$ are the PSF-corrected ellipticities of the galaxies which
serve as an unbiased estimator of the shear, and
$\e_{1,2}^{*,{\rm pol}}$ are the components of  
the PSF correction polynomial at the position $\fat{\theta}$ of a galaxy. 
We find that the correlation between stellar PSF and shear is greatly reduced
by the PSF correction, yet some non-zero residuals remain as we expected.
Another test for residual systematics from PSF correction in the final lensing 
catalogues is the $\map$ (defined in equation (\ref{map_q})) 
two-point cross-correlation function of uncorrected ellipticities of stars and
corrected ellipticities of galaxies (see Fig. \ref{mapcross}), 
\begin{equation}
\langle \map^* \map^{\rm gal} \rangle=\sum_{i=1}^{N_{\rm
    fields}}\langle \map^* \map^{\rm gal} \rangle_i\;.
\end{equation}
We find small non-zero residuals with different behaviour for the various 
survey parts. In particular, these residuals become increasingly non-zero with
growing aperture size for our own observations and part of the ASTROVIRTEL 
fields. We will come back to these two points later during our analysis in
Sect. \ref{shearselpeaks}, concluding that they do not affect our cluster
detection method in a noticeable way.

To summarise, our PSF correction effectively takes out mimicked coherent shear
patterns from the data, yet small residuals remain, which are much smaller 
than the coherent shear signal (a few percent) we expect from a typical
cluster at intermediate redshift range ($z=0.1\dots0.4$).

\subsection{Catalogue creation: object detection and calculation of lensing
  parameters}
Our catalogue production can be split into three parts, the object detection,
the calculation of the basic lensing quantities, and a final filtering of the
catalogue obtained. An absolute calibration of the magnitudes with high 
accuracy is not needed for this work since we are mostly interested in the 
shapes of the lensed galaxies but not in their fluxes. We adopted 
photometric standard zeropoints which bring us, conservatively estimated, to 
within $0.05-0.1$ mag of the real photometric zeropoint.

For the detection process we use \textit{SExtractor} \citep{bea96} to create a
primary source catalogue. The weight map created during the coaddition
 \citep[see][]{esd05} is used in this first step, guaranteeing that the highly
varying noise properties in the mosaic images are correctly taken into
account. This leads to a very clean source catalogue that is free from
spurious detections. The number of connected pixels (DETECT\_MINAREA) used in
this work for the object detection was 5, and we set the detection threshold
(DETECT\_THRESH) to 2.5. 
These thresholds are rather generous. $10-25\%$ of the objects detected
are rejected again in the later filtering process since they were too small
or too faint for a realiable shear measurement, or other problems appeared  
during the measurement of their shapes or positions.

In the second step we calculate the basic lensing quantities using
\textit{KSB} \citep{ksb95}. This includes PSF anisotropy correction, and the
recalculation of the objects' first brightness moments since
\textit{SExtractor} yielded positions with insufficient accuracy\footnote{This
  has been overcome in the recent \textit{SExtractor} releases (v2.4.3 and
  higher), using a Gaussian weighting function in the process.} for the
purpose of our analysis. Details of our implementation of
\textit{KSB} are given in \cite{ewb01}, and a mathematical description of the
PSF anisotropy correction process itself can be found in \cite{bas01}.

\subsection{Catalogue creation: filtering}
The third step consists of filtering the catalogue for various unwanted
effects. To avoid objects in the catalogue that are in the immediate vicinity
of bright stars, we explicitly set BACK\_TYPE = MANUAL and BACK\_VALUE = 0.0
(our coadded images are sky-subtracted) in the \textit{SExtractor}
configuration file. Thus \textit{SExtractor} is forced to assume a zero sky
background and does not model the haloes around bright stars as sky 
background. This proved to be a very efficient way of automatically masking 
brighter stars and the regions immediately surrounding them (see e.g. the
lower right panel of Fig. \ref{holes}). Further filtering on the
\textit{SExtractor} level is done by excluding all objects that are flagged
with FLAG $>4$ and those with negative half-light radii.

\begin{figure}[t]
\includegraphics[width=1.0\hsize]{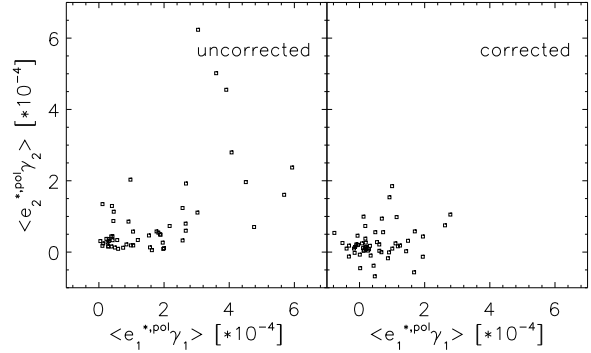}
\caption{\label{estar_gamma}{Shown is the correlation between stellar
    ellipticity and measured shear before (left panel) and after (right panel)
    PSF correction for the 58 survey fields. The median improvement is a
    factor of 3, but residuals are still present.}}
\end{figure}

\begin{figure}[t]
\includegraphics[width=1.0\hsize]{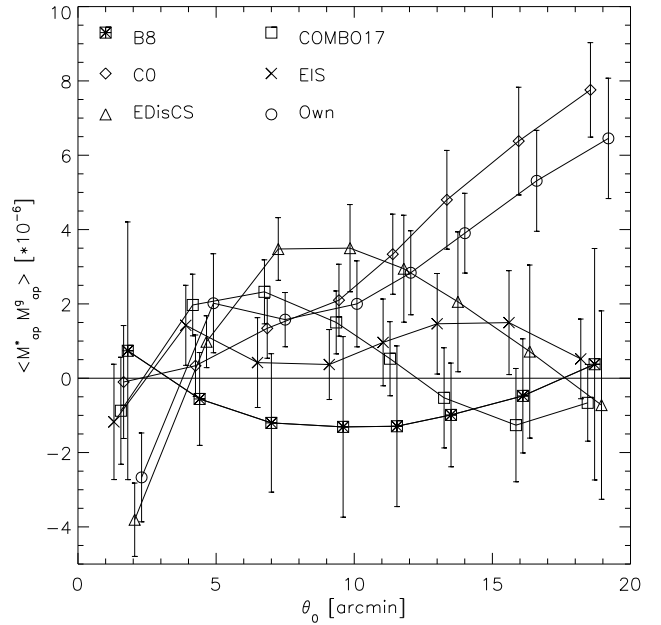}
\caption{\label{mapcross}{$\map$ two-point cross-correlation function of
    uncorrected ellipticities of stars and corrected ellipticities of 
    galaxies, binned for the various survey data sources. The analysis of the  
    ASTROVIRTEL fields has been split up into the B8- and C0-fields, since
    these exhibit different properties.}}
\end{figure}

On the \textit{KSB} level we filter such that only galaxies for which no 
problems in the determination of centroids occurred remain in the catalogue. 
Galaxies with 
half-light radii ($r_{\rm h}$) smaller than $0.1-0.2$ pixels than the left 
ridge of the stellar branch in an $r_{\rm h}$-mag-diagramme are 
rejected from the lensing catalogue, as are those with exceedingly bright 
magnitudes or a low detection significance ($\nu_{\rm max}<10$). See the left 
panel of Fig. \ref{rhmagnupg} for an illustration of these cuts. From the same
panel it can be seen that a significant number of galaxies have half-light
radii comparable to or a bit smaller than the PSF, which makes their shape
measurement noisier. Yet their number is large enough so that the shear 
selection of galaxy clusters profit significantly when these objects are 
included in the calculation. By including these objects, we gain $10-25\%$ in
terms of the number density of galaxies, and $3-10\%$ in terms of
signal-to-noise of the detections.

Furthermore, all galaxies with a PSF corrected modulus of the ellipticity 
larger than 1.5 are removed from the catalogue (the ellipticity can become
larger than 1 due to the PSF correction factors, but is then downweighted), as
are those for which the
correction factor $({\rm Tr}\,P^{\rm g})^{-1}>5$ \citep[see][]{ewb01}. The 
fraction of rejected galaxies due to the cut-off in $P^{\rm g}$ is relatively 
small, as can be seen from the right panel in Fig. \ref{rhmagnupg}. 

An overall impression of the remaining objects in the final catalogues is 
given in Fig. \ref{asteroid}. In total, typically $10-25$\% of the objects are
rejected from the initial catalogue due to the \textit{KSB} filtering steps.
The remaining average number density of galaxies per field is $n\sim 11\parcm$ 
(min: 6, max: 28), not corrected for the 
\textit{SExtractor}-masked areas (as described at the beginning of this
section; on the order of 5\% per field). For the width of the ellipticity 
distribution we 
measure $\sigma_\e=0.34$ for each of the two ellipticity components,
averaged over all survey fields. Both $n$ and $\sigma_\e$ determine
the signal-to-noise of the various mass concentrations detected.

\section{\label{chapter_map}Detection methods}
\subsection{\textit{S}-statistics and an optimal filter \textit{Q}}
We base our detection method on the aperture mass statistics ($\map$) 
as introduced by S96. $\map$ can be written as a filtered integral of the
tangential shear
$\gamma_{\rm t}$, 
\begin{equation}
\map(\fat\theta_0) = \int \td^2\varphi\;\gamma_{\rm t}
(\fat\varphi;\fat\theta_0)\;Q(\varphi)\,.
\label{map_q}
\end{equation}
Originally, the idea of $\map$ is to obtain a measure of the mass inside an
aperture independent of the mass sheet degeneracy in the weak lensing
case. Written in the form (\ref{map_q}) we can also simply interpret it as the
filtered amount of the tangential shear around a fiducial point $\fat\theta_0$
on the sky, where $\gamma_{\rm t}(\fat\varphi;\fat\theta_0)$ is the tangential
shear at position $\fat\varphi$ relative to $\fat\theta_0$, and $Q$ is some
radially symmetric spatial filter function. 

\begin{figure*}[t]
\includegraphics[width=1.0\hsize]{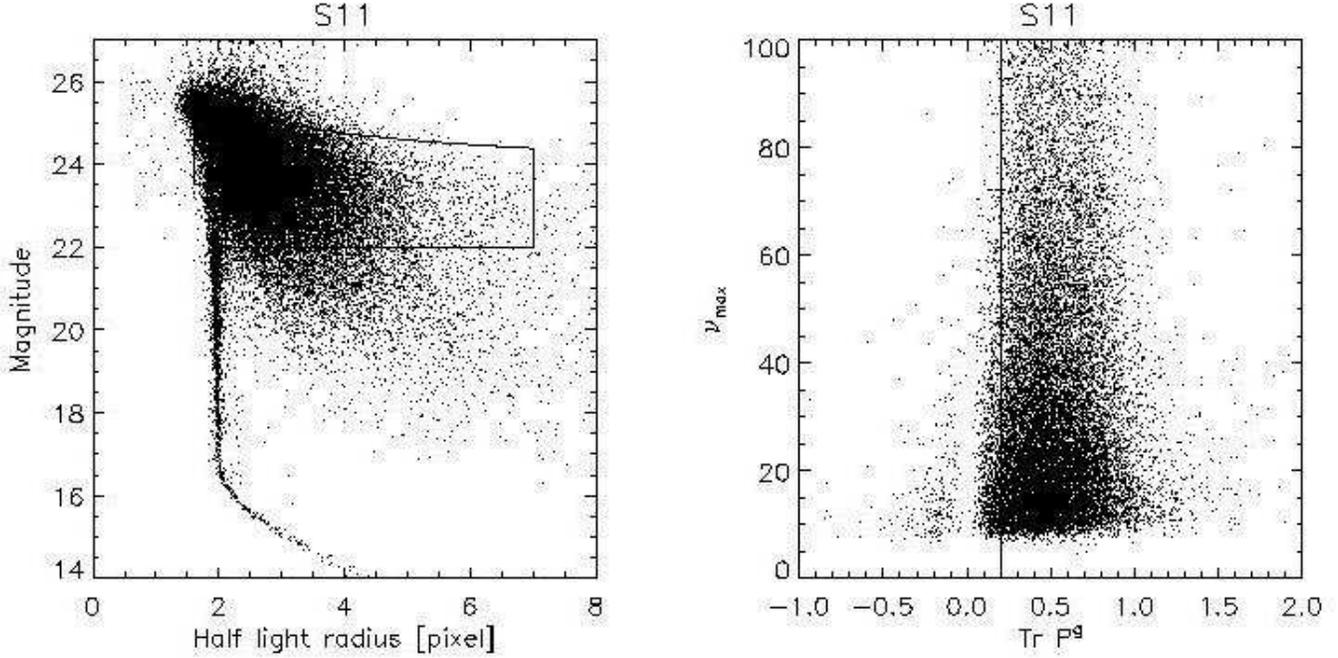}
\caption{\label{rhmagnupg}\small{Left panel: Stars appear as a vertical branch
  in a 
  $r_{\rm h}$-mag plot. Those brighter than $R=16.5$ saturate the detector
  and thus increase in size. The solid line encircles the galaxies which are 
  used for the lensing analysis. The upper curved line indicates a cut 
  in detection significance ($\nu_{\rm max}>10$), which has proven to work 
  significantly better than a constant cut on the faint end of the
  magnitudes. Right panel: $\nu_{\rm max}$ against $P^{\rm g}$. Objects left
  of the indicated threshold are rejected from the lensing
  catalogue. Typically 1\% of the galaxies are removed during this step.}}
\end{figure*}

The variance of $\map$ for the unlensed case, respectively the weak lensing 
regime, is given by
\begin{equation}
\sigma^2_{\map} = \frac{\pi\sigma_{\e}^2}n \int_0^{\theta}\td 
 \vartheta\, \vartheta\, Q^2(\vartheta)\,,
\end{equation}
where $\sigma_{\e}$ is the dispersion of intrinsic source
ellipticities, and $n$ the number density of background galaxies. The
integration is performed over a finite interval since we will have
$Q\sim 0$ for $\vartheta>\theta$, i.e. for radii larger than the aperture
size chosen. For the application to real data we will replace this integral in
Sect. \ref{discretecase} by a sum over individual galaxies.

We then define the \textit{S-statistics}, or the $S/N$ for $\map$
respectively the measured amount of tangential shear, as
\begin{equation}
S(\theta;\fat\theta_0) = \sqrt{\frac{n}{\pi\,\sigma_\e^2}}\;
   \frac{\int_0^\theta \td^2\vartheta\,
   \gamma_{\rm{t}}(\fat\vartheta;\fat\theta_0)\,Q(\vartheta)}
   {\sqrt{\int_0^{\theta}\td \vartheta\,\vartheta\, Q^2(\vartheta)}}\;.
\label{s_statistics}
\end{equation}
Here $\theta$ is the aperture radius, and $\vartheta$ measures the distance
inside this aperture from its centre at $\fat\theta_0$. This expression gets
a bit simplified due to assumed radial symmetry (see
Appendix \ref{sn_derivation}). Hereafter, we will call the 2-dimensional
graphical representation of the $S$-statistics the \textit{S-map}. If we plot
the $S$-statistics for a given mass concentration as a function of aperture
size, then we refer to this curve as the \textit{S-profile}.

The filter function $Q$ that maximizes $S$ for a given density (or shear)
profile of the lens can be derived using either a variational principle
\citep{sch04}, or the Cauchy-Schwarz inequality (S96). It is obtained for
\begin{equation}
Q(\vartheta)\propto\gamma_{\rm{t}}(\vartheta) \;,
\label{theor_optimalQ}
\end{equation}
where $\gamma_{\rm t}(\vartheta)$ is the tangential shear of the lens,
averaged over a circle of angular radius $\vartheta$. This is
intuitively clear, since a signal (i.e., $\gamma_{\rm{t}}$) with a certain
shape is best picked up by a similar filter function ($Q$). 
We present in Sect. \ref{filters} a mathematically simple family of filter
functions that effectively fulfills this criterion for the NFW mass profile.

\subsection{\label{discretecase}Formulation for discrete data fields}
For the application to real data, the previously introduced continuous 
formulation has to be discretised. First, we replace the tangential shear 
$\gamma_{\rm t}$ with the tangential ellipticity $\e_{\rm{t}}$, which 
is in the weak lensing case an unbiased estimator of $\gamma_{\rm t}$.
Thus we have
\begin{equation}
\map = \frac{1}{n}\sum_{i=1}^{N}\e_{{\rm t}i}\;Q_i\,,
\label{mapdiscrete}
\end{equation}
where $n$ is the number density of galaxies, and $N$
the total number of galaxies in the aperture. Introducing individual galaxy
weights $w_i$ as proposed by \cite{ewb01}, this becomes
\begin{equation}
\map = \frac{A}{\sum_i w_i}\;\sum_i\e_{{\rm t}i}\;
       w_i\;Q_i\,,
\label{mapdiscreteweight}
\end{equation}
where $A$ is the aperture area, previously absorbed in the number density
$n$. 

The noise of $\map$ can be estimated from $\map$ itself as was shown by 
\citet{krs99} and S96. In the weak lensing case its variance evaluates as
\begin{equation}
\sigma^2(\map) = \langle{\map}^2\rangle - \langle\map\rangle^2\ = 
\langle{\map}^2\rangle\;,
\end{equation}
since one expects $\langle\map\rangle=0$. Substituting with equation
(\ref{mapdiscrete}) yields 
\begin{eqnarray}
\nonumber
\sigma^2(\map) & = &
  \frac{1}{n^2}\;\sum_{i,j}\langle\e_{{\rm t}i}\,
  \e_{{\rm t}j}\rangle\,Q_i\,Q_j =
  \frac{1}{n^2}\;\sum_{i}\langle{\e_{{\rm t}i}}^2\rangle\,{Q_i}^2\\
  & = & \frac{1}{2n^2}\;\sum_{i} |\e_i|^2\,{Q_i}^2\,,
\end{eqnarray}
where we used 
\begin{equation}
\langle{\e_{{\rm t}}}^2\rangle = \frac{1}{2}\,|\e|^2\;,
\end{equation}
and the fact that $\e_{{\rm t}i}$ and $\e_{{\rm t}j}$ are 
mutually independent and thus average to zero for $i \ne j$. The $Q_i$ are 
constant for each galaxy and thus can be taken out of the averaging process.
Again, in the case of individual galaxy weights this becomes
\begin{equation}
\sigma^2(\map) = 
    \frac{A^2}{2\;\l(\sum_i w_i\r)^2}\;
    \sum_i|\e_i|^2\; w_i^2\;Q_i^2\,,
\end{equation}
using the fact that the $w_i$ are constant for each galaxy like the $Q_i$.
Therefore, we obtain for the $S$-statistics
\begin{equation}
S = \frac{\sqrt{2}\; \sum_i \e_{{\rm t}i}\,w_i\,Q_i}
  {\sqrt{\sum_i |\e_i|^2\,w_i^2\,{Q_i}^2}}\;,
\label{sn_analytic}
\end{equation}
which grows like $\sqrt{N}$.

\subsection{Validity of the $\fat\map$ concept for our data}
If $\map$ is evaluated close to the border of an image or on a data field with
swiss-cheese topology due to the masking of bright stars, then the aperture 
covers an `incomplete' data field. Therefore, the returned value of $\map$
does no longer give a result in the sense of its original definition in S96,
which was a measure related to the filtered surface mass density inside the
aperture. Yet it is still a valid measure of the tangential shear inside the
aperture, including the $S/N$-estimate in equation (\ref{sn_analytic}), and
can thus be used for the detection of mass concentrations.

Since the number density of background galaxies inside an aperture is
not a constant over the field due to the masking of brighter stars (and the
presence of the field border), we have to check for possible unwanted
effects. As long as the holes in the galaxy distribution are small compared to 
the aperture size, and as long as their number density is small enough so that 
no significant overlapping of holes takes place, the effects on the 
$S$-statistics are negligible (see \figref{holes}). In fact, the decreased
number density just leads to a lowered significance of the peaks detected in
such areas, without introducing systematic effects.

If the size of the holes becomes comparable to the aperture, spurious peaks
appear in the $S$-map at the position of the holes. This is because the
underlying galaxy population changes significantly when the aperture is moved
to a neighbouring grid point. When such affected areas were present in our
data, then we excluded them from the statistics and masked them in the 
$S$-maps, even though these spurious peaks are typically not very significant 
($\sim 2\sigma$). Our threshold for not evaluating the $S$-statistics at a 
given grid point is reached if the effective number density of the galaxies in
the aperture affected is reduced by more than 50\% due to the presence of
holes (or the image border). Spurious peaks become very noticeable if the
holes cover about 80\% of the aperture. This is rarely the case for our data
unless the aperture size is rather small (2\myarcminnodot), or a particular
star is very bright. We conclude that our final statistics is free from any
such effects.

\subsection{\label{filters}Filter functions}
As expressed in (\ref{theor_optimalQ}), an optimal filter function should
resemble the tangential shear profile. In the following such a filter $Q(x)$ 
is constructed, assuming that the azimuthally averaged shear dependence is 
caused by an NFW density profile \citep{nfw97} of the lensing mass
concentration. We set $x:=\vartheta/\theta$, with $x$ being the projected
angular separation $\vartheta$ on the sky from the aperture centre, in units
of the aperture radius $\theta$. By varying $\theta$, shear patterns
respectively mass concentrations of different extent can be detected.

\citet{wrb00} and \citet{bar96} derived an expression for the tangential shear 
of the universal NFW profile. Based on their finding we can construct a new
filter function $Q_{\rm NFW}$ over the interval $x\in [0,1]$, having the shape
\begin{equation}
Q_{\rm NFW}(x) = \;\;\;\;\;\;\;\;\;\;\;\;\;\;\;\;\;\;\;\;\;\;\;\;\;\;\;
\label{q_nfw}
\end{equation}
\begin{equation}
\nonumber
\begin{cases}
  \frac{4(3y^2-2)}{y^2(y^2-1)\sqrt{1-y^2}}\;
  {\rm{arctanh}}\sqrt{\frac{1-y}{1+y}}+\frac{4}{y^2}\,
  {\rm{ln}}\frac{y}{2}+\frac{2}{1-y^2}
        & \;\;(y < 1) \\[0.3cm]
  \frac{10}{3}-4\,{\rm{ln}}\,2
        & \;\;(y = 1) \\[0.15cm]
  \frac{4(3y^2-2)}{y^2(y^2-1)\sqrt{y^2-1}}\;
  {\rm{arctan}}\sqrt{\frac{y-1}{1+y}}+\frac{4}{y^2}\,
  {\rm{ln}}\frac{y}{2}+\frac{2}{1-y^2}
        & \;\;(y > 1)
 \end{cases}\;
\end{equation}
Here we defined $y:=x/x_{\rm c}$, with $x_{\rm c}$ being a dimensionless
parameter changing the width (and thus the sharpness) of the filter over the
interval $x\in[0,1]$, in the sense that more weight is put to smaller radii
for smaller values of $x_{\rm c}$\footnote{Thus $x_{\rm c}$ is in analogy to
  the NFW scale radius $r_{\rm s}$}. This expression is smooth and
continuous for $y=1$, and approaches zero as ${\rm ln}(y)/y^2$ for $y\gg 1$. 

\begin{figure}[t]
\includegraphics[width=1.0\hsize]{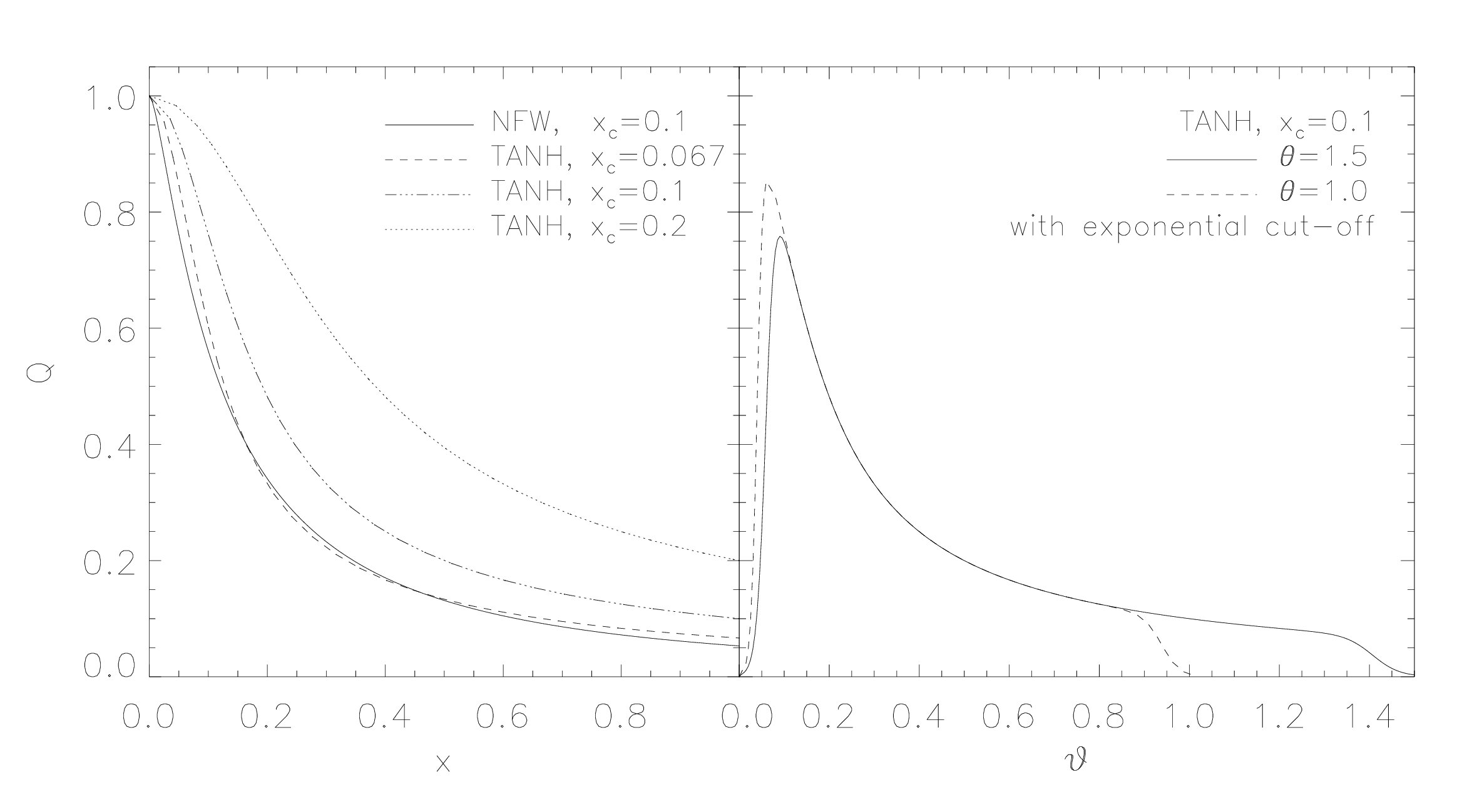}
\caption{\label{nfwtanh}\small{Left panel: $Q_{\rm NFW}$ (solid line),
  shown with some representations of $Q_{\rm TANH}$. The exponential cut-off 
  $E(x)$ for small and large radii is not introduced in this plot in order to
  show the differences between the two filter types better. As can be seen,
  the $Q_{\rm TANH}$ filter is a very good approximation for $Q_{\rm NFW}$, 
  giving slightly more weight to smaller radii. Right panel: $Q_{\rm TANH}$
  with the cut-off introduced by $E(x)$ at both ends, plotted against the
  radial coordinate $\vartheta$. The dashed line ($\theta=1$) can be compared 
  directly to the dash-dotted line in the left panel.}}
\end{figure}

Due to the mathematical complexity of $Q_{\rm NFW}$, the calculation of the
$S$-statistics is rather time consuming for a field with $\sim10^4$
galaxies. We introduce an approximating filter 
function with simpler mathematical form that produces similarly good results 
as $Q_{\rm NFW}$. It is given by
\begin{equation}
Q_{\rm TANH}(x) = E(x)\;\frac{{\rm tanh}\,(x/x_{\rm c})}{x/x_{\rm c}},
\label{qtanhmod}
\end{equation}
having the $1/x$ dependence of a singular isothermal sphere for 
$x\gg x_{\rm c}$. The hyperbolic 
function, having a $\propto x$ dependence for small $x$, absorbs the 
singularity at $x=0$ and approaches 1 for growing values of $x$. The
pre-factor $E(x)$ is a box filter, independent of $x_{\rm c}$, with
exponentially smoothed edges. It reads
\begin{equation}
E(x) = \frac{1}{1+{\rm e}^{6\,-\,150 x}+{\rm e}^{-47\,+\,50 x}}
\end{equation}
and lets $Q$ drop to zero for the innermost and outermost 10\% of the 
aperture, while not affecting the large rest (see right panel
of Fig. \ref{nfwtanh}). It is introduced because both $Q_{\rm NFW}$ and 
$Q_{\rm TANH}$ (without this pre-factor) are not zero at the centre nor at 
the edge of the aperture. This cut-off is very similar to that introduced in 
[S96] for a different radial filter profile. It suppresses stronger
fluctuations when galaxies  
enter (or leave) the aperture, receiving significant non-zero weight. It also
avoids assigning a large weight to a few galaxies at the aperture centre, 
which as well can lead to significant fluctuations in the $S$-statistics when 
evaluated on a grid. The effect of $E(x)$ is rather mild though, since usually 
several hundred galaxies are covered by one aperture unless it is of very
small size so that $E(x)$ becomes important.

We thus have a filter function based upon the two-dimensional parameter space
$(\theta\,,x_{\rm c})$. The differences between $Q_{\rm TANH}$ and 
$Q_{\rm NFW}$ are indistinguishable in the noise once applied to real data,
so that we do not consider $Q_{\rm NFW}$ henceforth.

It is not the first time that $\map$ filters following the tangential shear
profile are proposed or used. We have already utilised the filter in equation
(\ref{qtanhmod}) to confirm a series of luminosity-selected galaxy clusters
\citep{ses04}. Before that, \citet{psp03} approximated $Q_{\rm NFW}$
with
\begin{equation}
Q_{\rm PAD}(x) = \frac{2\;{\rm ln}(1+x)}{x^2}- \frac{2}{x(1+x)}-1/(1+x)^2\;,
\label{qpad}
\end{equation}
which was later-on modified by \citet{hes05}. They multiplied 
(\ref{qpad}) with a 
Gaussian of certain scale radius in order to suppress the effects of the 
cosmic shear that become dominant for larger radii. Even though the
mathematical descriptions are different, the latter two filters are in effect
very similar to (\ref{qtanhmod}), and we could not find one of them superiour
over the other based on our rather inhomogeneous data set. Hence, we do not
consider them for the rest of the work. The validity of our approach has
recently been confirmed by \citet{msm06}, who also use a filter following the
tangential shear profile, individually adapted for each field to
minimise the effect of cosmic shear. Based on the same data as
we use in the present paper, they find that our filter defined in equation
\ref{qtanhmod} yields very similar results as compared to their optimised
filter, which means that the lensing effects of the large scale structure in
our rather shallow survey are not very dominant.

Differences in the efficiency of such ``tangential'' filters are thus expected
to arise for very deep surveys only, and/or in case of high redshift clusters
($z=0.6$ and more, for which our survey is not sensitive). In all other cases
they are hardly distinguishable from each other since the noise in the images
and the deviations from the assumed radial symmetry of the shear field and the
NFW profile are dominant. Thus we consider the $Q_{\rm TANH}$ filter to be
optimally suited for our survey. For a comparison with other filters that do
not follow the tangential profile, see the example shown in \figref{s11_a901}.

\begin{figure}[t]
\center{\includegraphics[width=1.0\hsize]{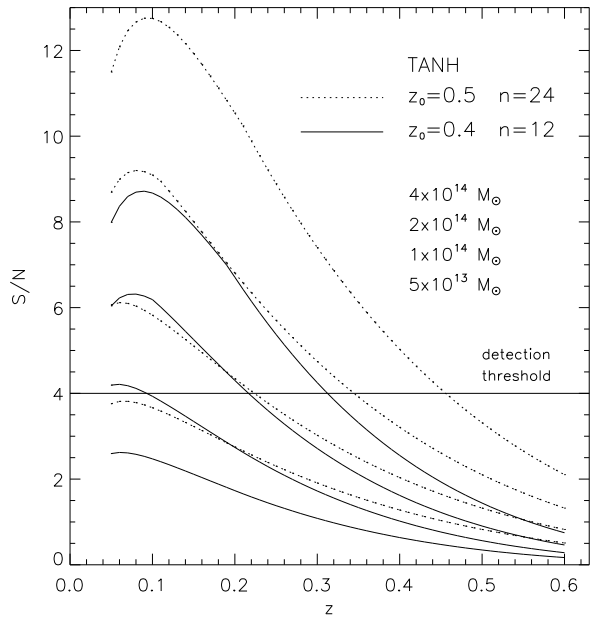}}
\caption{\label{sntanh}Expected \textit{optimal} S/N ratios for NFW dark
  matter haloes 
  for four different cluster masses ($M_{200}$) and two different image
  depths. The mathematical derivation of the $S/N$ for a particular cluster at
  a given redshift is given in Appendix \ref{sn_derivation}. Note that the
  filter scale is not constant along each of the curves since the shear fields
  get smaller in angular size with increasing lens redshift. Note also that we
  used a lower integration limit of $z_{\rm d}=0.2$ for the lensed background
  galaxies (see Sect. \ref{sensitivity} for both aspects).} 
\end{figure}

\subsection{\label{sensitivity}Sensitivity}
Figure \ref{sntanh} gives an idea of the sensitivity of our selection method
for the $Q_{\rm TANH}$ filter (see Appendix \ref{sn_derivation} for
mathematical details). 
This plot was calculated taking into account various characteristics of our
survey and analysis. First, we introduced a maximum aperture radius of 
20\myarcminnodot to reflect the finite field of view of our fields. This
affects (lowers) the $S/N$ of very low redshift ($z<0.09$) clusters, whose 
shear fields can be very extended, and leads to a more distinct maximum of the 
curves. Second, the aperture size is not a constant along these curves but 
limited by the angular distance where the strength of the tangential shear 
drops below 1\%, merging into the cosmic shear. Lastly, we used 
a lower redshift cut-off of $z_{\rm d}=0.2$ in order to roughly reflect our 
selection criteria for the background galaxies (see Fig. \ref{rhmagnupg}). The 
effect of the latter cut is minor, it increases the $S/N$ on the order of 5\% 
for the low redshift range we probe, as compared to no cut.

As a result, our $S$-statistics is insensitive for structures of mass equal or 
less
than $M_{200}=5\times10^{13}$ M$_\odot$ for all redshifts. In data of average
depth ($n=12\parcm$) we can detect mass concentractions of
$1,2,4\times10^{14}$ M$_\odot$ out to $z=0.10, 0.22$ and 0.32,
respectively. The same objects would still be seen at $z=0.22, 0.34$ and 0.46
in the deeper exposures with twice the number density of usable background
galaxies.

\subsection{\label{pstatistics}Introducing the $P$-statistics}
In addition to the $S$-statistics defined above, we introduce a new measure
which we call \textit{peak position probability statistics}, hereafter
simply referred to as \textit{P}-statistics. It tells us if at one particular
position on the sky the $S$-statistics makes significantly more detections for
various filter scales than one would expect in the absence of a lensing
signal. As we find in Sect. \ref{shearselpeaks}, the $S$- and $P$-statistics
complement each other rather well.

\begin{figure*}[t]
\includegraphics[width=1.0\hsize]{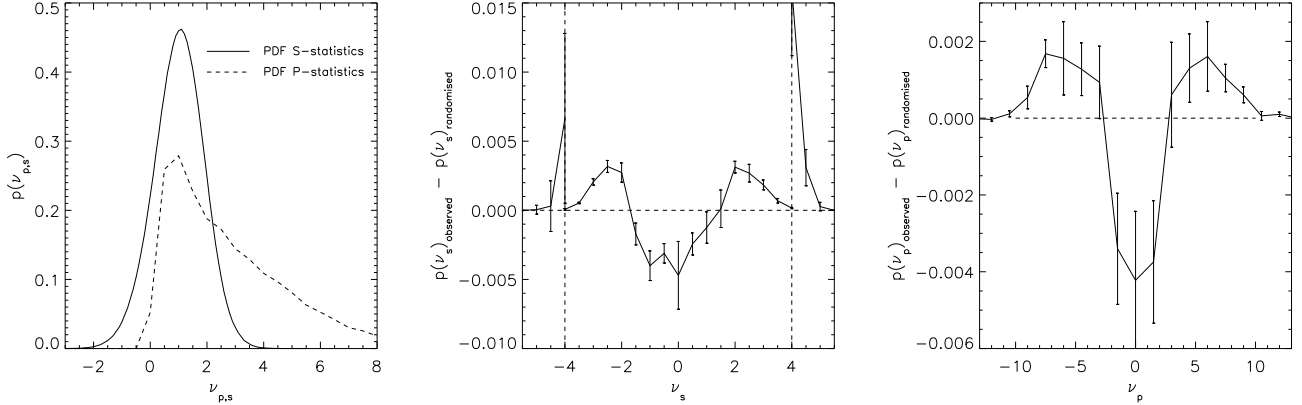}
\caption{\label{pdf_sstat}Left panel: The normalised PDF for the peaks based
  on the entire survey area, averaged over all scales and $x_{\rm c}$. Whereas 
  the peak PDF for the $S$-statistics can 
  be well approximated by a Gaussian, the PDF for the $P$-statistics is highly 
  non-Gaussian with a broad tail. The $\nu_p$ should therefore not be directly 
  interpreted as $S/N$. The middle ($S$-statistics) and right panel
  ($P$-statistics) show the difference between the observed peak PDF and the 
  PDF obtained from 10 randomisations. The middle plot is magnified by 
  $\times 100$ for $|\nu_s|\geq 4$ for better visualisation. Both these plots
  also contain the minimum (and mostly negative) peaks, i.e. underdense 
  regions or voids, which accounts for the symmetric appearance. A significant 
  excess of peaks and voids exists in the observed data (as compared to the
  randomised data) for $|\nu_s|>2$ and $|\nu_p|>3$. The error bars are taken
  from the randomisations.}
\end{figure*}

We calculate the $P$-statistics as follows:
\begin{itemize}
\item{For each given survey field we look up the positions of all peaks
  detected in the full $(\theta\,,x_{\rm c})$ parameter space.}
\item{We then select all positive peaks with a signal-to-noise
  larger than 2.5, and map their positions on the sky. The individual
  positions are weighted with the corresponding peak $S/N$. This discrete map 
  is then smoothed with a Gaussian kernel of width
  40\myarcsecnodot, yielding a continuous map. The smoothing is introduced to
  smooth out the variations in the positions determined for the very same peak
  using different filter scales. In other words, it links two detections made
  nearby on the sky in different filters to the same mass concentration.}
\item{We randomise the orientations of the galaxies in the original catalogue
  10 times and repeat the above step. From that we derive a constant noise
  level for each survey field. The assumption of a constant noise is an
  approximation which we based on a set of 200 randomisations of an arbitrary
  survey field. We find that the noise is constant to within 15\% or less.
  It can be well estimated with a set of just 10 randomisations.}
\item{The ``true'' map is then divided by the rms obtained from the noise map, 
  yielding an estimate of how reliable the peak at the given position is. We 
  then call this rescaled map the $P$-map in analogy to the $S$-map. However,
  note that the $P$-statistics has a very non-Gaussian probability 
  distribution function (see also the left panel of Fig. \ref{pdf_sstat}).}
\end{itemize}

The main idea behind this approach is that a real peak has an extended shear
field, i.e. it will be picked up by a larger number of different filter 
scales. In other words, as the aperture size changes, different samples
of galaxies are used and all of them will yield a signal above the detection
threshold (provided a sufficient lensing strength). On the contrary, a
spurious peak mimicked by the noise of the intrinsic galaxy ellipticity is not
expected to show such a behaviour, thus the $P$-statistics will prefer a true
peak over a false peak. It has an advantage over the $S$-statistics since it
looks at a broader range of filter scales instead of one single scale. Thus it
is capable of giving significance to a peak that otherwise goes unnoticed by
the $S$-statistics. On the other hand, objects with weak shear fields will
not be recognised by the $P$-statistics since they appear only for one or a
few nearby filter scales.

We would like to emphasise that we introduced the $P$-statistics for this work
on an experimental basis only. Its performance has not yet been evaluated
based on simulations, but it yields very similar results as the
$S$-statistics (see Sect. \ref{shearselpeaks}), so that we included it in 
this presentation.

There is some arbitrariness in the way we implemented the $P$-statistics. For
example, the lower threshold of 2.5$\sigma$ for the peaks considered can be
decreased
or increased. The former would make it smoother since more peaks are
included, but does not yield any further advantage since it picks up too much
noise. Increasing the threshold beyond 3.5 reduces the number of peaks
entering the statistics significantly. This makes the determination of the
noise level unstable, and one starts losing less significant peaks.
There is also room for optimisation
concerning the selection of input data, as for this work we simply included
peaks from the full parameter space. Concentrating on a smaller set of filter
scales could yield a more discriminative power, but bears the risk of losing 
objects. In addition, the smoothing length has been chosen to obtain the best 
compromise between smoothing out position variations in the lensing
detections while maintaining the spatial resolving power of the
$P$-statistics. The chosen kernel width of 40\myarcsecnodot appears optimal 
for our survey, but may well be different for other data sets.

In order to distinguish between individual $S/N$ measurements made with the 
$P$- and $S$-statistics, we will use the terms \nup and \nus henceforth.

\subsection{Validation of the $S$- and $P$-statistics}
As a consistency check for the concept of the $S$- and the $P$-statistics, one 
can compare the peak probability distribution function (PDF) of the observed 
data against randomised data sets. The presence of cluster lensing should 
distort the PDF in the sense that more peaks are detected for higher $S/N$ 
values \citep[see][for an example]{mhs02}.

To this end we created 10 copies of our entire survey catalogue with
randomised galaxy orientations, destroying any lensing signal, but keeping the 
galaxy positions and thus all other data characteristics fixed. We then
calculated the PDFs for all local maxima (overdense regions) and 
minima (underdense regions) of the
observations and the randomisations, accumulating the detections from the 
entire parameter space probed. The middle and right panel of Fig.
\ref{pdf_sstat} show the difference between the PDFs of the observed and the
randomised data sets. Both PDFs, for the $S$- and the $P$-statistics, are 
significantly skewed, showing an excess of peaks and voids above thresholds 
of about $\nu_s>2$ and $\nu_p>3$.

\section{\label{shearselpeaks}Shear-selected mass concentrations}
\subsection{Selecting the cluster candidates}
The way we established our sample of possible mass concentrations (``peaks'') 
is as follows. The $S$-statistics is evaluated
\begin{itemize}
\item{on a grid with a 10\myarcsecnodot spacing}
\item{for 19 different filter scales\footnote{1.6, 2.0, 2.4, 2.8, 3.2, 3.6,
    4.0, 4.8, 5.6, 6.3, 7.1, 7.9, 8.7, 9.9, 11.9, 13.9, 15.9, 17.9, 19.8
    arcminutes. The odd numbers arise from the original definition which was
    made in units of pixels.}.}
\item{for 7 different scale parameters $x_{\rm c}$\footnote{0.025, 0.05, 0.1,
    0.2, 0.5, 1.0, 2.0}}
\item{From these $S$-maps (133 different representations per survey field) 
    we include all peaks higher than 4$\sigma$, which corresponds to the lower 
    limit of the $(S/N)_{\rm min}=4\dots 5$ range that is considered useful by 
    various authors \citep[see][for example]{reb99, hty04}.}
\end{itemize}

The $P$-statistics is evaluated on the same grid space and obtained as
described in Sect. \ref{pstatistics}. We will use as well a detection 
threshold of $\nu_p = 4.0\sigma$, which has shown to provide us with a similar
number of detections as the $S$-statistics, so we can compare the two samples.

The detections made with both statistics are summarised in Tables
\ref{shearselclusters1} to \ref{shearselclusters3}. Those peaks seen with the 
$S$-statistics have the best matching filter scale reported, i.e. the one
yielding the highest $S/N$. Column 1 contains numeric labels for the
detections, followed by a string indicating with which statistics it was
made. The third and fourth column contain the detection significances \nup and
\nus (if applicaple). Columns five, six and seven carry a classification
parameter (see 
below), a richness estimate of a possible optical counterpart if present, plus
the distance of the peak from the latter. This is followed by the filter scale
and the $x_{\mathrm{c}}$ parameter (if applicable), and then the name of the
survey field in which the detection was made. Finally, we report the redshift
of an optical counterpart if known.

\subsection{\label{defclass}Classification of the peaks detected}
In order to characterise the line of sight for a peak in terms of visible
matter, we introduce a rough classification scheme based on the $R$-band 
images. We visually inspected a radius of 2\myarcmin0 around the peak
position for apparent overdensities of brighter galaxies, as compared to the
surrounding field. Even though this is a rather crude approach, it is
good enough to tell if a peak is likely associated with some luminous 
matter, or not. The radius of 2\myarcmin0 has been chosen since we observed
from known galaxy clusters that the lensing detection can scatter up to
1\myarcmin5 with respect to the optical center of the cluster (see
Table \ref{positionoffsets}). This is either due to substructure in the
cluster, or due to noise (the detection is made from a finite number of
galaxies only).

The classes are defined as follows and are based on galaxies taken from the 
range $R\sim 17-22$ (thus they are not member of the catalogue of 
background galaxies, see also the left panel of Fig. \ref{rhmagnupg}):

\begin{itemize}
\item{class 1: concentration of more than 50 galaxies}
\item{class 2: 35-50 galaxies} 
\item{class 3: 25-35 galaxies} 
\item{class 4: 15-25 galaxies} 
\item{class 5: 5-15 galaxies} 
\item{class 6: no counterpart visible} 
\end{itemize}

Judging from our number counts \citep[][]{hed06} for the ESO Deep Public
Survey data, we expect a number density of about 11000 galaxies with $R\leq22$
per square degree. Assuming a random distribution of these foreground galaxies
and neglecting clustering effects, we obtain a scattering of $\sigma=7.2$ for 
the total number of galaxies within the 2 arcmin radius. Thus, class 5 objects
represent not more than a 2.1$\sigma$ overdensity as compared to the
randomised distribution. 

We consider classes 1 to 4 to be reliable optical counterparts, and refer to
them as \textit{bright peaks} henceforth. Classes 5 are rather dubious, 
and go as \textit{dark peaks} together with those lines of sight classified 
as 6. See Fig. \ref{classes} for an illustration of bright peaks of classes
$1-4$.

The boundaries between the classes are permeable. For example, if we find
an overdensity of 12 galaxies, and 4 or 5 of them stand out from the rest 
by their brightness and are of elliptical type, the class 5 object would 
become a class 4. Similarly, if we find 20 galaxies of similar brightness, 
but they show a significantly higher concentration than the rest of the 
sample, it becomes class 3. On the other hand, if the distance between
the mass peak and the center of the optical peak exceeds 100 arcseconds, 
we decrease the class by one step. The same holds if the galaxies seen appear 
to be at redshift higher than $\sim 0.3$ or more, then we lower the rank
by 1 since our selection method becomes less sensitive with increasing 
redshift. About 20\% of our sample were up- or downgraded in this way.

\subsection{Spectroscopically ``confirmed'' candidates}
For 22 out of the 72 bright peaks, we found spectra in the  
\textit{NASA Extragalactic
  Database}\footnote{http://ned.ipac.caltech.edu}. Most of them come from the  
\textit{SDSS}\footnote{http://www.sdss.org} or the 
\textit{Las Campanas Distant Cluster Survey} \citep{gzd01}. 
For 7 cases we have only two spectra, thus they just indicate a 
cluster or group nature, but we can not take it as hard 
evidence. Those redshifts are marked with an asterisk in Tables 
\ref{shearselclusters1} to \ref{shearselclusters3}. For 15 other peaks
the cluster nature was already known or has meanwhile been secured, 
either by spectroscopy, photometric redshifts, or by other photometric 
means \citep[see e.g.][for the red sequence method]{gly00}. 
Three of them (\#043, 053 and 157) turn out to be clusters or groups
of galaxies at different redshifts projected on top of each other,
with the previous two being triple. For simplicity, we refer to all 
these objects in the following as ``confirmed'' peaks, even though 
more spectroscopic data has to be obtained for most of them for 
sufficient evidence. 

Whenever spectra were available, they confirmed our assumption of spatial
concentrations in all cases. In order to secure the 50 most promising
candidates, we recently started a large spectroscopic survey aiming at between
20 and 50 galaxies per target. This will not only pin down the redshifts
of the possible clusters, but also allow us to identify further projection 
effects and in some cases possible physical connections with nearby peaks
(e.g. \#029, 056, 074, 084, 092, 128, 136, 141, 158). We will report these
results in future papers.

The $1\sigma$ redshift range of the peaks confirmed so far is 
$z = 0.09\dots0.31$. We therefore predominantly probe the lower redshift range
of clusters, which is consistent with the theoretically expected sensitivity
of our survey \citep[see][and \figref{sntanh}]{krs99}. 

\begin{table}[t]
   \caption{\label{positionoffsets}Average angular offsets between the peak 
     and the optical counterpart}
   \begin{tabular}{lccc}
   \noalign{\smallskip}
   \noalign{\smallskip}
   \hline 
   \hline 
   \noalign{\smallskip}
   \noalign{\smallskip}
   class & number & average & \\
         & of peaks & offset [ \myarcsecnodot]& \\
   \noalign{\smallskip}
   \noalign{\smallskip}
   \hline 
   \noalign{\smallskip}
   1 &  3 & $53\pm38$ & \\
   2 & 14 & $59\pm24$ & \\
   3 & 21 & $51\pm16$ & \\
   4 & 34 & $55\pm30$ & \\
   5 & 43 & $62\pm32$ & \\
   6 & 43 & N/A & \\
   \hline 
   \end{tabular}
   \normalsize
\end{table}

\subsection{Positional offsets}
Table \ref{positionoffsets} shows that the peaks coincide with the positions
of the optical counterparts to within 0\myarcmin9$\pm$0\myarcmin5,
independent of the peak classification. On the one hand, these offsets are due 
to noise, since the shear field is obtained by a finite number of lensed 
galaxies with intrinsic ellipticities. In addition, in general the shear 
fields deviate from the radial symmetry assumed by the $\map$ filter. 
On the other hand, these offsets can be physical in the sense that light does
not trace mass very well for young or still non-virialised clusters.
The two largest clusters in our sample, Abell 901 (\#039) and Abell 1364
(\#082), are good examples. For them, the positions of the weak lensing
detections are shifted away from the cD galaxies in the direction of
sub-clumps. See also Fig. \ref{classes} for an illustration.

\begin{figure}[t]
\includegraphics[width=1.0\hsize]{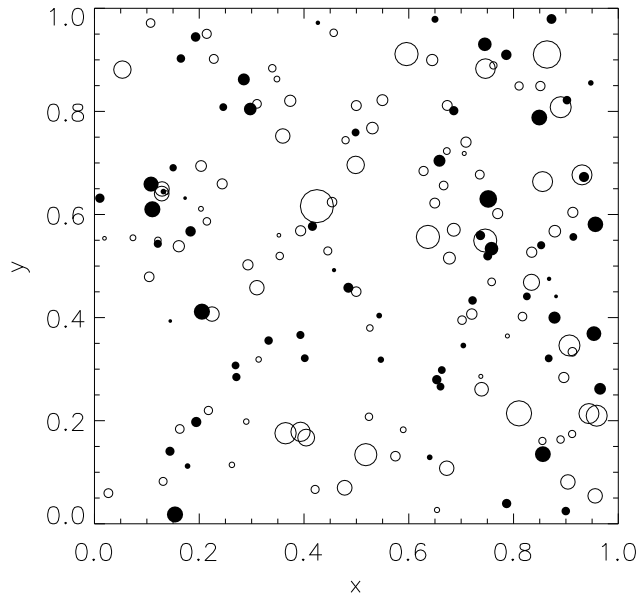}
\caption{\label{spatialdist}{Shown is the spatial distribution of the 158
    detected peaks for the WFI field of view. Open symbols indicate the bright
    peaks with classes 1-4, the filled ones peaks with classes 5 and 6. The 
    symbol size represents the detection significance. The pattern is
    indistuinguishable from a random distribution, and we also do not see
    differences for peaks obtained with either the $S$- or the
    $P$-statistics (not shown).}}
\end{figure}

\begin{figure}[t]
\includegraphics[width=1.0\hsize]{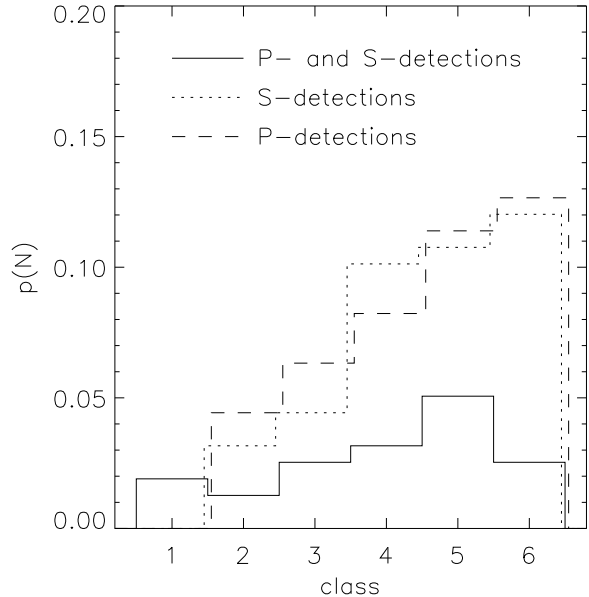}
\caption{\label{p_and_s_classes}{Shows the fraction of detections
    made in either the $S$- or the $P$-statistics, or in both.}}
\end{figure}

\subsection{\label{comparing_SP} Comparing the $S$- and $P$-statistics}
We make 90 and 95 detections with the $S$- and the $P$-statistics,
respectively, having 26 peaks in common. As can be seen from
Table \ref{fractionbrightdark}, the number of bright and dark peaks
is equally balanced between the two methods, both yielding slightly 
more dark than bright peaks. A few fields exist in which either only 
the $S$-statistics (e.g. CL1216-1201) or the $P$-statistics (e.g. 
FIELD17\_P3) make detections, but the individual number of detections 
made per field are in general small. The statistics do not show a preference
with respect to particular fields. The same holds for the spatial
distribution of peaks as a function of position in the detector 
mosaic (see also Fig. \ref{spatialdist}), or the occurrence of dark 
peaks as a function of exposure time (see Fig. 
\ref{fraction_brightdarkexptime} for a merged plot of the two statistics). 
In general, the overlap between the two statistics increases if the 
detection thresholds of $S/N=4$ are further lowered, i.e. peaks  
seen in only one of the statistics then appear also in the other. 
Yet then the overall number of detections increases to many 
hundreds, and we did not evaluate the overlapping fraction in this 
case. 

The only noteworth difference between the $S$- and the $P$-statistics 
is that the latter detects about 30\% peaks of classes 2 and 3, whereas 
the $S$-statistics returns 25\% more peaks of class 4. The occurence of dark
peaks is indistuinguishable between the two methods (see
Fig. \ref{p_and_s_classes}).

\subsection{Detector and survey field biases, dark peaks}
To check for systematics concerning the WFI@2.2 detector array, we plotted the 
positions of all peaks with respect to the array geometry (Fig. 
\ref{spatialdist}). It appears that the right half is a bit more crowded than 
the left part of the detector array. After running a dozen random 
distributions with the same number of objects we find that this is 
insignificant. Thus the distribution is random-like for both bright and dark 
peaks, and does not prefer or avoid particular regions.

Upon counting the bright and dark peaks in the five main data sources of our
survey (see Sect. \ref{surveysources}), we find differences in the ratio
between bright and dark peaks (Table \ref{surveydiffs}). Namely, the
ASTROVIRTEL and EIS data, and our own observations, show an excess of
$20-50\%$ in terms of dark peaks as compared to the bright peaks, and are
about comparable to each other. The EDisCS
survey has a factor 2.1 more dark peaks, but is also that part of our survey 
with the most shallow exposure times. Contrary, the COMBO-17 data has twice as
many bright as darks peaks, but this comes not as a surprise since the S11
and A901 fields are centered on known galaxy clusters with significant
sub-structure. If we subtract the known clusters and all detections likely
associated with them, we still have an ``excess'' of 40\% for the bright peaks
in COMBO-17. Again, this is not unplausible since the COMBO-17 fields form by
far the deepest part of our survey, which let us detect more mass
concentrations. But this holds for both bright \textit{and} dark peaks, as the 
number of detections per square degree shows (Table \ref{surveydiffs}).

\begin{figure}[t]
\includegraphics[width=1.0\hsize]{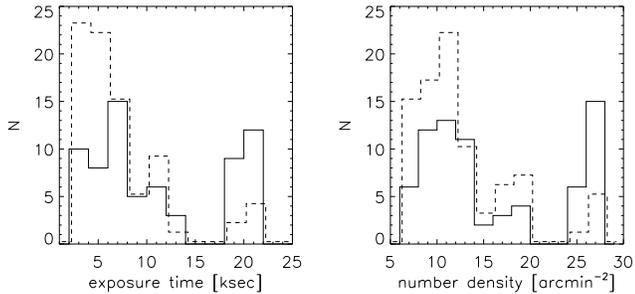}
\caption{\label{fraction_brightdarkexptime}Shown is the number of bright peaks
  (solid line) and dark peaks (dashed line) as a function of exposure time 
  (left) and galaxy number density (right). Shallow exposures with low number
  density have more dark peaks than deeper exposures.}
\end{figure}

In order to check if the dark peaks might arise due to imperfect PSF
correction, we compare their occurrence with the remaining PSF residuals
in our lensing catalogues (Figs. \ref{estar_gamma} and \ref{mapcross}).
Therein we do not find evidence that the imperfect PSF correction
gives rise to dark peaks. However, 
Fig. \ref{fraction_brightdarkexptime} indicates that small exposure times
(less than $10-12$ ksec) and/or a low number density (less than 
$n\sim13-15\parcm$) of galaxies foster the occurrence of dark peaks. Yet this
has to be seen with some caution, in particular because we have only a small
number of deep fields (mainly COMBO-17) as compared to the shallow ones, and
the deep fields are prtially concentrated on known structures. If we take the
known 
structures into account and remove them from the statistics, we are still left
with a smaller fraction of dark peaks in the deep exposures, but the question
remains in how far the particular pointings of those fields introduce a bias. 
To answer this question empirically, we would need about 10 empty fields of 20
ksec exposure time each.

Due to the $\map$ filter we use, and to the large inhomogeneity of our survey,
we can not directly compare the occurrence of dark peaks in our data to 
existing numerical simulations. Also, these simulations usually make
significantly more optimistic assumptions in terms of usable number 
density of galaxies and field of view than we could realise with GaBoDS
\citep[see][for example]{reb99,jvw00,hes05}. In particular \citet{hty04}
have shown that in their simulations ($n=30$) they expect to detect 43 
real peaks (efficiency of $\sim 60\%$), scaled to the same area as GaBoDS and
drawn with $S/N>4$ from mass reconstruction maps. A similar number of false 
peaks appear as well, being either pure noise peaks, or peaks with an 
expected $S/N<4$ being pushed over this detection limit. The latter 
would be labelled as bright peaks in our case. Our absolute numbers 
are different (72 bright and 86 dark peaks) since we use a very different
selection method. Yet, if we interpret our dark peaks as noise peaks,
the ratio between our bright and dark peaks is comparable to the ratio between
their true and false peaks.

This interpretation, i.e. dark peaks are mostly noise peaks, is strengthened
by the fact that with increasing peak $S/N$ the fraction of dark peaks is
decreasing (see Table \ref{fractionbrightdark}), for both the $S$- and the
$P$-statistics. However, our observational data base (sky coverage) is too
small to tell if this trend, i.e. the dark peaks dying out, continues for
higher values of the $S/N$.

\begin{figure}[t]
\includegraphics[width=1.0\hsize]{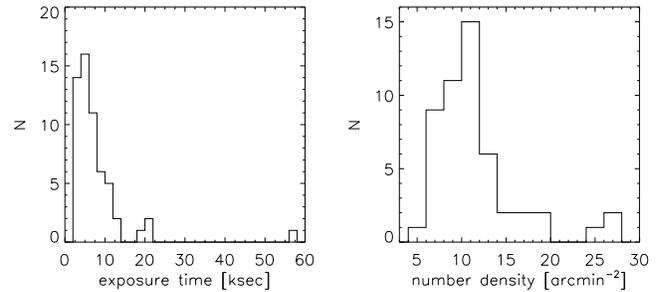}
\caption{\label{gabodsexptime}{Left: Histogram of the exposure times. 
    The peak at 57
    ksec represents the Chandra Deep Field South (CDF-S). Right: Number
    density of galaxies in the 58 fields after all filtering steps, leaving a
    total number of about 710000 usable galaxies. The distribution reflects
    the distribution of exposure times shown in the left panel.}}
\end{figure}

\section{Summary}
In the present paper we have introduced a new type of filter function for the
aperture mass statistics. This filter function follows the tangential shear
profile created by a radial symmetric NFW density profile. We have shown
that it is optimally suited for an application to our 19 square degree weak
lensing survey conducted with the WFI@2.2m MPG/ESO telescope. This is a
survey with very inhomogeneous depth, in which we expect to be able to
detect mass concentrations in the redshift range $z=0.1\dots0.5$. 
We also introduced the new $P$-statistics, currently on a test-basis only.
It turned out to deliver very similar results in the number of bright and dark
peaks detected, and for the time being looks like a good complement to the
$S$-statistics. Its performance has to be investigated more deeply though,
and there is space for optimisation. The global PDFs for both the $S$- and
$P$-statistics show clear excess peaks for higher values of $S/N$ as compared
to the randomisations. Thus the presence of lensing mass concentrations in 
our survey data is confirmed.

We introduced a classification scheme in order to associate the hypothetical
mass peaks detected with possible luminous matter. From the 158 detections 
we made with the combined $S$- and $P$-statistics, 72 (46\%) appear to have 
an optical counterpart. For 22 of those we found spectra in the literature,
confirming the above mentioned redshift range, and that indeed a mass
concentration exists along those particular lines of sight. For a smaller
number of those we have spectroscopic evidence that the peaks detected are due
to projection effects. We expect that in our currently conducted spectroscopic
follow-up survey more such projection cases will be uncovered, together with a
confirmation of a very significant fraction of the remaining bright peaks.
In a future paper we will also compare this shear-selected sample with an 
optically selected sample using matched filter techniques.

We gained some insight into the subject of dark peaks. They appear
preferentially in shallow data with a small number density of galaxies,
indicating that a large fraction of those could be due to noise (i.e. 
instrinsic galaxy ellipticies), or they are statistical flukes. Nevertheless, 
in our deep fields we also observe a significant fraction of dark peaks, but
our statistics can not be unambiguously interpreted since  
those fields are biased towards clusters with significant sub-structure, and
we have only a very small number of them. Nevertheless, real physical objects
such as very underluminous clusters are far from being ruled out at this
point. At least on the mass scale of galaxies the last year has seen
astonishing examples of objects having several $10^8\;M_\odot$ of neutral 
hydrogen, yet apparently no star formation has taken place in them
\citep{mrd05, aul06}. Whether similar objects can still exist on the cluster
mass scale is unclear. 

Finally, we would like to repeat that the Garching-Bonn Deep Survey has been
made with a 2m telescope. Most numerical simulations done so far are much more
optimistic in terms of the number density reached ($n\sim 30\parcm$) and 
correspond to surveys that are currently conducted (or will be in the near
future) with 4m- and 8m-class telescopes, such as SUPRIME33 \citep{mhr05} or
the CFHTLS\footnote{http://www.cfht.hawaii.edu/Science/CFHLS/}. One noteworthy 
exception will be the
KIDS\footnote{http://www.strw.leidenuniv.nl/$\sim$kuijken/KIDS/} 
(\textit{Kilo Degree Survey}) obtained with OmegaCAM@VST, covering 1500 square
degrees starting in 2007.

\begin{acknowledgements}
This work was supported by the BMBF through the DLR under the project 50 OR 
0106, by the BMBF through DESY under the project 05AE2PDA/8, and by the 
DFG under the projects SCHN 342/3--1 and ER 327/2--1. Furthermore we
appreciate the support given by ASTROVIRTEL, a project funded by the European
Commission under FP5 Contract No. HPRI-CT-1999-00081. The authors thank
Ludovic van Waerbeke at UBC for very fruitful discussions.
\end{acknowledgements}
\bibliography{cluster}

\appendix

\section{\label{sn_derivation}Expected S/N for a NFW halo} 
We derive the $S/N$ expected for a radially symmetric NFW dark matter halo,
using a flat cosmology with $\Omega_0=0.3$, $\Omega_\lambda=0.7$, $h=0.7$ and 
$\sigma_8=0.85$.

According to S96, the S/N for $\map$ for a cluster at the 
origin of the coordinate system evaluates as
\begin{eqnarray}
S(\theta) & = & \sqrt{\frac{4 \pi n}{\sigma_\e^2}}\;
   \frac{\int_0^\theta \td \vartheta\, \vartheta\,
   \gamma_{\rm{t}}(\vartheta)\,Q(\vartheta)}
   {\sqrt{\int_0^{\theta}\td \vartheta\,\vartheta\, Q^2(\vartheta)}}\;.
\label{appdx1}
\end{eqnarray}
Here we assumed radial symmetry (yielding a factor of $2 \pi$), write $\theta$ 
for the aperture radius and $\vartheta$ for the distance from the centre of 
the aperture. $n$ is the number density of galaxies, and $\sigma_\e$ is the
dispersion of the modulus of the galaxy ellipticities.

The tangential shear for a radially symmetric NFW profile was given by 
\citet{wrb00} as
\begin{equation}
  \gamma_{\rm{t}}(x) = \frac{r_s \delta_c \rho_c}{\Sigma_{\rm{cr}}} \;g(x)
  \label{nfwtangshear}
\end{equation}
where
\begin{eqnarray}
  g(x) & = & \begin{cases}
    g_<(x) & \;\;(x < 1) \\[0.3cm]
    \frac{10}{3}-4\,{\rm{ln}}\,2 & \;\;(x = 1) \\[0.15cm]
    g_>(x) & \;\;(x > 1)
  \end{cases}\\
  x & = & \frac{\Dd \; \vartheta}{r_s}\\
  \rho_c & = & \frac{3}{8 \pi G}\;H_0^2\;[\Omega_0 (1+z)^3 + \Omega_\Lambda]\\
  \Sigma_{\rm{cr}}&=&\frac{c_L^2}{4 \pi G}\;
  \frac{1}{\Dd}\l<\frac{\Dds}{\Ds}\r>^{-1}\;.
\end{eqnarray}
Here, $c_L$ is the speed of light and $G$ is Newton's constant. The NFW
concentration parameter $c$ is a function of cosmology, the mass of the
cluster and its redshift. We calculate $c$ using the \textit{cens} routine
kindly provided by J. 
Navarro\footnote{http://www.astro.uvic.ca/$\sim$jfn/cens/}. $r_s$ and 
$\delta_c$ are the NFW scale radius and characteristic over-density of the
cluster. $\Sigma_{\rm{cr}}$ is the \textit{critical surface mass density},
where we write $\Dd$, $\Ds$ and $\Dds$ for the angular diameter distances
between the observer and the lens, between the observer and the source, and
between the lens and the source, respectively. For a flat cosmology, they are
defined as 
\begin{equation}
  D(z_1,z_2) = \frac{c_L}{H_0}\;\frac{1}{1+z_2}
    \int_{a_2}^{a_1} {\td a \l[a\;\Omega_0 + a^4\;\Omega_\Lambda\r]^{-1/2}}\;,
\end{equation}
with $z_1<z_2$ and $a=1/(1+z)$. To take into account the redshift distribution
of the lensed galaxies, we assume that those follow the normalised
distribution given in \citet{bbs96}, with parameterisation
\begin{equation}
  p(z) = \frac{3}{2 z_0}\;\l(\frac{z}{z_0}\r)^2\;{\rm{exp}}
         \l[-\l(\frac{z}{z_0}\r)^{3/2}\r]\;.
\label{redshift_distribution}
\end{equation}
The ratio $\Dds/\Ds$ is averaged over this redshift distribution, starting
with the lens redshift $z_{\rm{d}}$ as a lower integration limit.
The latter was chosen because galaxies with $z<z_{\rm{d}}$ are unlensed
and largely removed from our catalogues by appropriate detection thresholds
and cut-offs\footnote{The exact effect of this filtering on the shape of
the redshift distribution is not yet investigated. It will mostly affect
the $S/N$ prediction of high-redshift clusters, and only very little the
low-z regime probed with our data.}.

The functional expression for $g(x)$ is identical to the one already given
in equation (\ref{q_nfw}), and contains the shape of the shear profile. 
Finally, fixing the remaining numerical parameters provides us with all 
information to calculate the $S/N$. From our data we have $\sigma_\e=0.48$, 
and we assume two different image depths which we base on empirical
findings. One is shallow with $n=12\parcm$ and $z_0=0.4$, and the deeper one
given by $n=24\parcm$ and $z_0=0.5$.

The $S/N$ then evaluates as
\begin{equation}
  S(\theta) = \sqrt{\frac{4 \pi n}{\sigma_\e^2}}\; 
              \frac{r_s \delta_c \rho_c}{\Sigma_{\rm{cr}}}\;
   \frac{\int_0^\theta \td \vartheta\, \vartheta\,
   g(\vartheta)\,Q(\vartheta)}
   {\sqrt{\int_0^{\theta}\td \vartheta\,\vartheta\, Q^2(\vartheta)}}\;.
\end{equation}
\clearpage
\section{Further tables and figures}

\begin{table}[ht]
   \caption{\label{gabodsfields}The 58 fields used for this work} 
   \tiny
   \begin{tabular}{lrrrl}
   \noalign{\smallskip}
   \noalign{\smallskip}
   \hline 
   \hline 
   \noalign{\smallskip}
   Field & $\alpha$(2000.0) & $\delta$(2000.0) & $T_{\rm{exp}}$ & Source\\
   \noalign{\smallskip}
   \hline 
   \noalign{\smallskip}
   A1347\_P1    & 175.257 & $-$25.514 & 13500 & own observation\\
   A1347\_P2    & 175.792 & $-$25.509 & 7500  & own observation\\
   A1347\_P3    & 175.239 & $-$25.009 & 7000  & own observation\\
   A1347\_P4    & 175.794 & $-$24.998 & 8000  & own observation\\
   A901         & 149.077 & $-$10.027 & 18100 & COMBO-17\\
   AM1          & 58.811  & $-$49.667 & 7500  & ASTROVIRTEL\\
   B8m1         & 340.348 & $-$10.089 & 4500  & ASTROVIRTEL\\
   B8m2         & 340.348 & $-$10.589 & 5400  & ASTROVIRTEL\\
   B8m3         & 340.346 & $-$11.088 & 5400  & ASTROVIRTEL\\
   B8p0         & 340.348 & $-$9.590  & 7200  & ASTROVIRTEL\\
   \noalign{\smallskip}
   \hline 
   \noalign{\smallskip}
   B8p1         & 340.346 & $-$9.089  & 4500  & ASTROVIRTEL\\
   B8p2         & 340.345 & $-$8.589  & 5400  & ASTROVIRTEL\\
   B8p3         & 340.345 & $-$8.089  & 5400  & ASTROVIRTEL\\
   C0400        & 214.360 & $-$12.253 & 4800  & ASTROVIRTEL\\
   C04m1        & 214.727 & $-$12.753 & 4000  & ASTROVIRTEL\\
   C04m2        & 214.478 & $-$13.253 & 4000  & ASTROVIRTEL\\
   C04m3        & 215.318 & $-$13.753 & 4000  & ASTROVIRTEL\\
   C04m4        & 215.111 & $-$14.253 & 4000  & ASTROVIRTEL\\
   C04p1        & 214.726 & $-$11.753 & 4000  & ASTROVIRTEL\\
   C04p2        & 214.727 & $-$11.253 & 4000  & ASTROVIRTEL\\
   \noalign{\smallskip}
   \hline 
   \noalign{\smallskip}
   C04p3        & 215.098 & $-$10.753 & 4000  & ASTROVIRTEL\\
   CAPO-DF      & 186.037 & $-$13.107 & 13000 & ASTROVIRTEL\\
   CDF-S        & 53.133  & $-$27.822 & 57200 & GOODS EIS\\
                &            &             &       & COMBO-17\\
   CL1037$-$1243  & 159.444 & $-$12.754 & 3600  & EDisCS\\
   CL1040$-$1155  & 160.139 & $-$11.963 & 3600  & EDisCS\\
   CL1054$-$1146  & 163.581 & $-$11.813 & 3600  & EDisCS\\
   CL1054$-$1245  & 163.647 & $-$12.797 & 3600  & EDisCS\\
   CL1059$-$1253  & 164.755 & $-$12.920 & 3000  & EDisCS\\
   CL1119$-$1129  & 169.784 & $-$11.525 & 3600  & EDisCS\\
   CL1138$-$1133  & 174.508 & $-$11.599 & 3600  & EDisCS\\
   \noalign{\smallskip}
   \hline 
   \noalign{\smallskip}
   CL1202$-$1224  & 180.645 & $-$12.441 & 3600  & EDisCS\\
   CL1216$-$1201  & 184.170 & $-$12.062 & 3600  & EDisCS\\
   CL1301$-$1139  & 195.467 & $-$11.630 & 3600  & EDisCS\\
   CL1353$-$1137  & 208.306 & $-$11.598 & 3600  & EDisCS\\
   CL1420$-$1236  & 215.066 & $-$12.649 & 3600  & EDisCS\\
   Comp           & 65.307  & $-$36.283 & 5300  & ASTROVIRTEL\\
   DEEP1a       & 343.795 & $-$40.198 & 7200  & EIS\\
   DEEP1c       & 342.328 & $-$40.207 & 3900  & EIS\\
   DEEP1e       & 341.966 & $-$39.528 & 9000  & EIS\\
   DEEP2a       & 54.372  & $-$27.815 & 6000  & EIS\\
   \noalign{\smallskip}
   \hline 
   \noalign{\smallskip}
   DEEP2d       & 52.506  & $-$27.817 & 3000  & EIS\\
   DEEP2e       & 53.122  & $-$27.304 & 7500  & own observation\\
   DEEP2f       & 53.669  & $-$27.324 & 7000  & own observation\\
   DEEP3a       & 171.245 & $-$21.682 & 7200  & EIS\\
   DEEP3b       & 170.661 & $-$21.709 & 9300  & EIS\\
   DEEP3c       & 170.019 & $-$21.699 & 9000  & EIS\\
   DEEP3d       & 169.428 & $-$21.701 & 9300  & EIS\\
   FDF          & 16.445  & $-$25.857 & 11840 & COMBO-17\\
   F17\_P1      & 216.419 & $-$34.694 & 10000 & own observation\\
   F17\_P3      & 217.026 & $-$34.694 & 10000 & own observation\\
   \noalign{\smallskip}
   \hline 
   \noalign{\smallskip}
   F4\_P1       & 321.656 & $-$40.251 & 9500  & own observation\\
   F4\_P2       & 321.719 & $-$39.767 & 7000  & own observation\\
   F4\_P3       & 322.320 & $-$40.237 & 10000 & own observation\\
   F4\_P4       & 322.323 & $-$39.726 & 7500  & own observation\\
   Pal3         & 151.432 & $-$0.003  & 4200  & ASTROVIRTEL\\
   S11          & 175.748 & $-$1.734  & 21500 & COMBO-17\\
   SGP          & 11.498  & $-$29.610 & 20000 & COMBO-17\\
   SHARC-2      & 76.333  & $-$28.818 & 11400 & own observation\\
   \noalign{\smallskip}
   \hline 
   \end{tabular}
   \normalsize
\end{table}

\begin{figure*}[t]
\includegraphics[width=1.0\hsize]{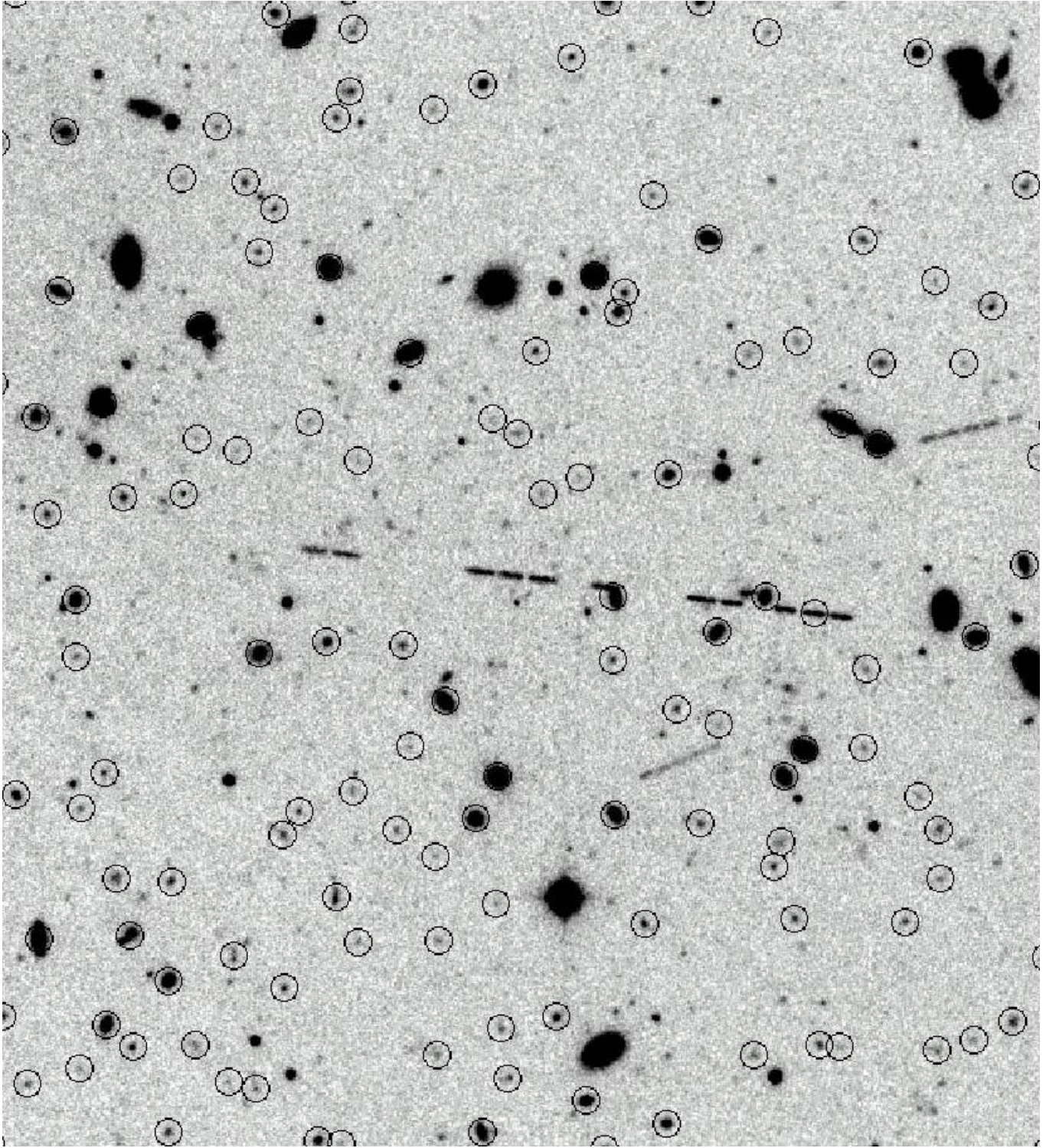}
\caption{\label{asteroid}\small{Indicated are the objects that are left over
  in the lensing catalogue after all typical filtering
  steps. Only fainter background galaxies are kept. Brighter sources, spurious
  detections, stars and highly elliptical objects such as asteroid tracks are 
  largely absent from the catalogue. The field of view is about 
  3\myarcminnodot.}}
\end{figure*}

\begin{figure*}[t]
\includegraphics[width=1.0\hsize]{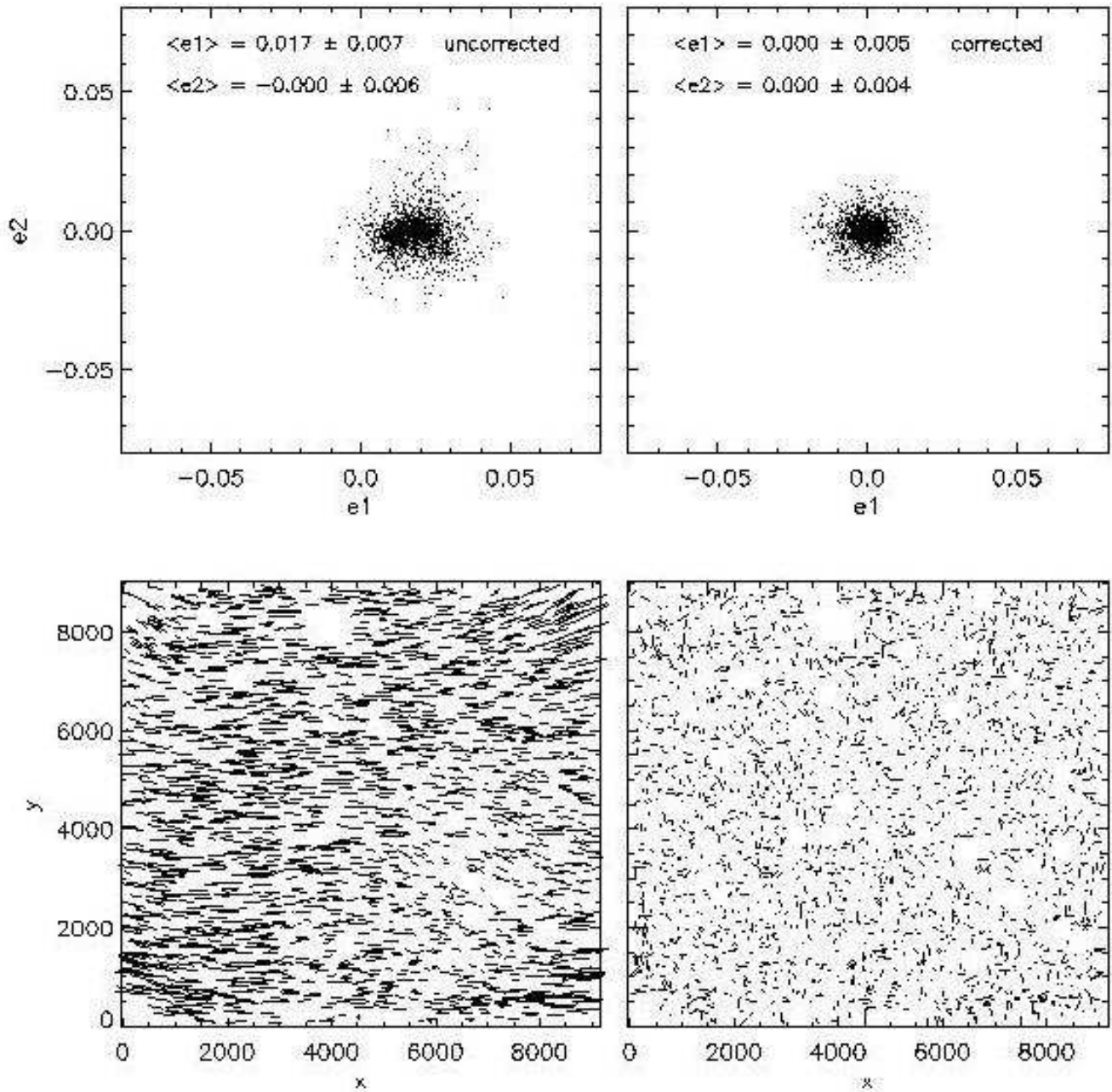}
\caption{\label{mosaniso_2}{Typical PSF anisotropy. Upper row: $\e_{1,2}$ 
scatter plot for stars before and after PSF correction. Lower left: PSF 
anisotropy pattern before correction. This plot is a more intuitive 
representation of the left panel above. Lower right: PSF residuals after
a polynomial fit was used to correct for the anisotropy. No coherent shear
signal is left.}}
\end{figure*}

\begin{figure*}
\includegraphics[width=1.0\hsize]{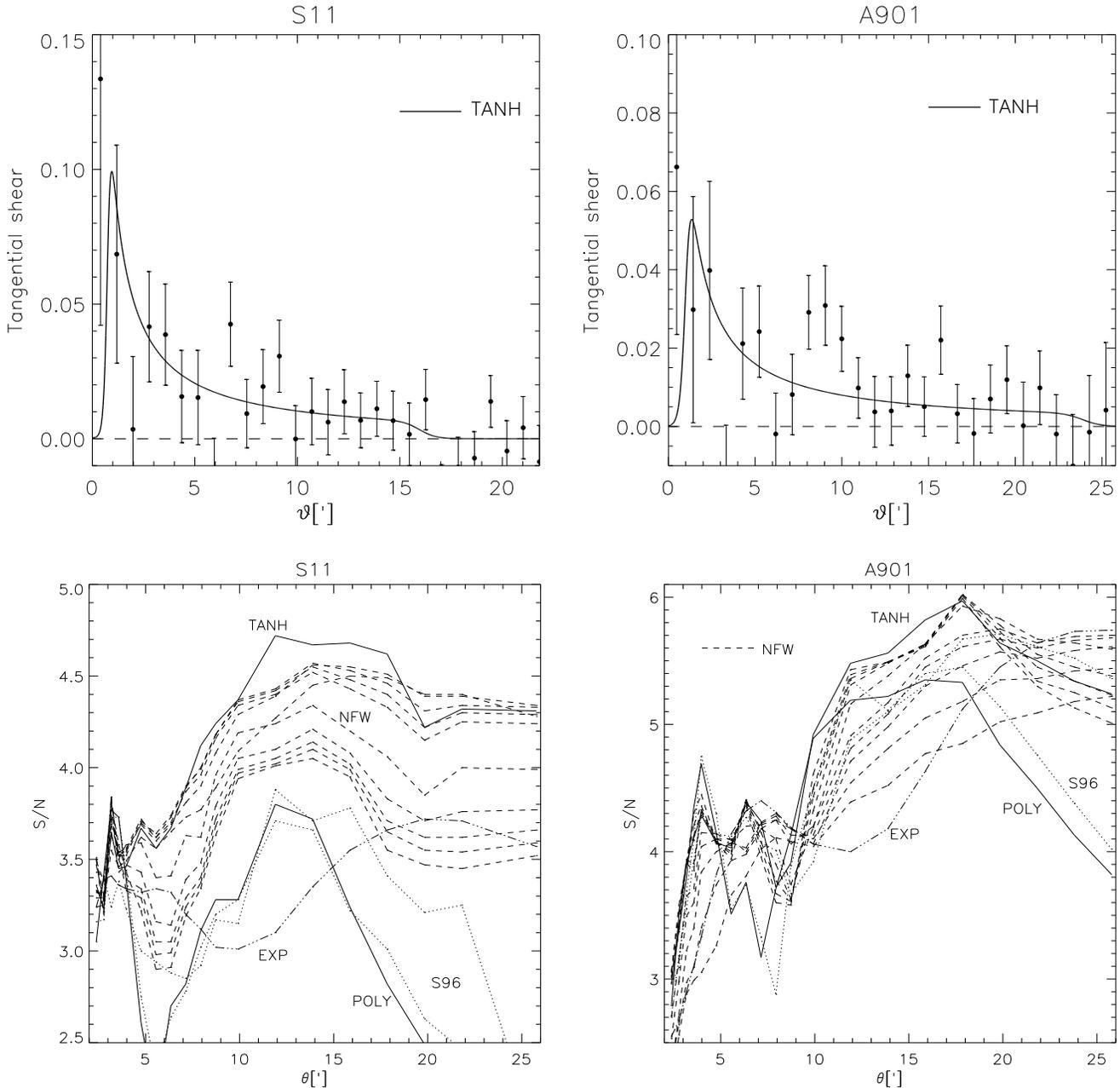}
\caption{\label{s11_a901}\small{Tangential shear and $S$-profile for the two
  largest clusters in the survey data. In the upper row the tangential shear
  is shown, with the TANH filter that yielded the highest $S/N$ overlaid as a 
  solid line. For better comparison the amplitude of TANH was scaled so 
  that it best fits the tangential shear. All data points are mutually 
  independent. The lower row shows the $S$-profile of the two clusters for
  different filters $Q$. The NFW filter is plotted for 10 different values of  
  $x_{\rm c} \in [0.01,1.5]$, to show the scatter delivered by $x_{\rm c}$.
  The filters defined by \citet{psp03} and \citet{hes05} (not shown) deliver 
  very similar results.
  For comparison we plot various other types of filter functions introduced
  in the literature. POLY is the polynomial filter defined by \citet{svj98},
  mainly as a new measure for cosmic shear rather than cluster detection. The 
  filters S96 were defined by \citet{sch96_2}.
  EXP is based on the difference of two Gaussians of different width
  \citep{sch04}. Clearly, all of them yield smaller $S/N$ values than those
  following the tangential shear profile. If one of those filters yields
  a higher $S/N$ for a cluster than TANH in our data, then this is marked
  accordingly in Tables \ref{shearselclusters1} and \ref{shearselclusters2}.
  \newline
  Note: Although the 
  tangential shear is smaller for A901 than for S11, the $S/N$ is higher due 
  to the larger number density of galaxies with measured shapes ($n=15$ 
  arcmin$^{-2}$ for S11, and $n=24$ arcmin$^{-2}$ for A901).}}
\end{figure*}

\begin{figure*}[h]
\center{\includegraphics[width=1.0\hsize]{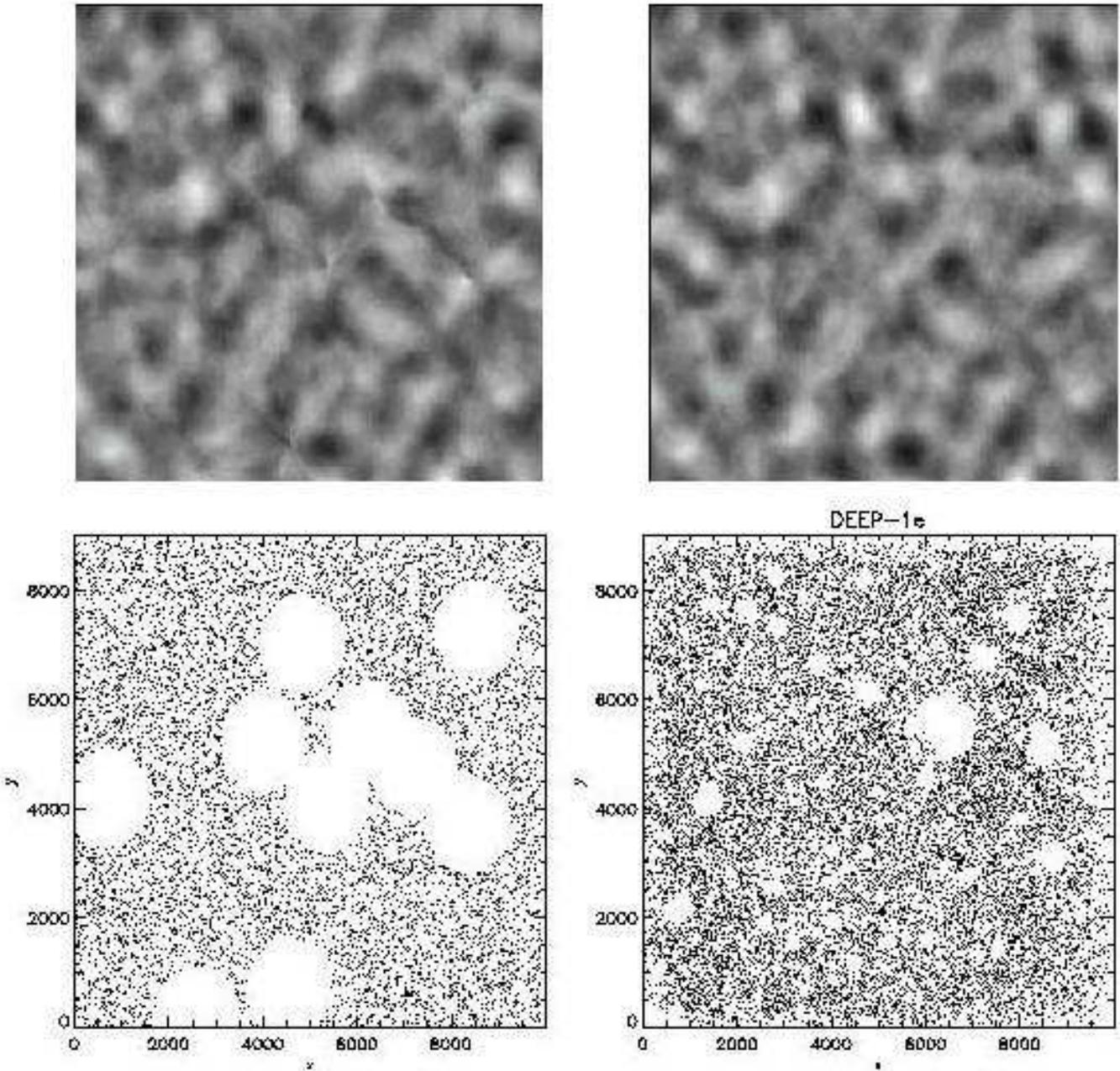}}
\caption{\label{holes}\small{Upper right: $S$-map for 
  randomised galaxy orientations and positions. Upper left: The same $S$-map, 
  evaluated after 10 randomly positioned holes with a radius of 90\% of the
  filter scale were cut into the data field (lower left). Artificial features
  show up in the $S$-map at the positions of the holes, the latter being 
  exaggerated in number and size for better visualisation. Lower right: True
  galaxy distribution of the field from which the galaxies were drawn for this
  test. The largest hole is caused by an 8th magnitude star. We conclude that 
  holes in the data fields are in general not a cause of concern for our 
  analysis.}}
\end{figure*}

\begin{table*}[t]
   \caption{\label{shearselclusters1}Shear-selected mass concentrations (part
     1). The first column contains a label for the peak, and the second one
   indicates if the detection was made with either the $S$- or the
   $P$-statistics (or both). The next two columns contain the corresponding 
   significances.  The classification shows if along the line of sight an
   overdensity of galaxies is found, with class 1 meaning a very obvious
   overdensity, and class 6 no overdensity. The richness indicates how many
   galaxies were found as compared to the average density in this field, and
   we give the distance between the peak and the optical counterpart. Finally, 
   we have the name of the survey field in which the detection was made, and
   possibly a redshift for the counterpart. An asterisk behind the redshifts
   indicates that it is based on the measurement of less than three galaxies.}
   \begin{tabular}{lrrrrrrrrll}
   \noalign{\smallskip}
   \noalign{\smallskip}
   \hline 
   \hline 
   \noalign{\smallskip}
   \noalign{\smallskip}
   Peak & stat & \nus & \nup & class & rich & dist & scale & 
   $x_{\mathrm{c}}$ & Field & z \\
   \noalign{\smallskip}
   \noalign{\smallskip}
   \hline 
   \noalign{\smallskip}
   \#001 & S  & 4.40 &      & 4 & 30 & 90 & 2\myarcmin8 & 2.0 & SGP &  \\
   \#002 & S  & 4.09 &      & 4 & 20 & 90 & 2\myarcmin4 & 2.0 & SGP &  \\
   \#003 & S  & 4.17 &      & 3 & 25 & 90 & 3\myarcmin6 & 2.0  & SGP &  \\
   \#004 & SP & 5.01 & 4.26 & 1 & 40 & 10 & 19\myarcmin8 & 0.025 & SGP & 0.257 \\
   \#005 & S  & 4.31 &      & 2 & 35 & 60 & 6\myarcmin3 & 0.05 & SGP & 0.5 \\
   \#006 & S  & 4.13 &      & 5 & 30 & 40 & 5\myarcmin6 & 0.025 & SGP &  \\
   \#007 & S  & 4.15 &      & 5 & 50 & 0  & 5\myarcmin6 & 0.025 & SGP &  \\
   \#008 & S  & 4.51 &      & 6 &    &    & 7\myarcmin9 & 2.0 & SGP &  \\
   \#009 & S  & 4.22 &      & 2 & 40 & 20 & 3\myarcmin6 & 0.1 & SGP &  \\
   \#010 & S  & 4.17 &      & 6 &    &    & 2\myarcmin0 & 2.0 & FDF &  \\
   \hline 
   \#011 & S  & 4.09 &      & 5 & 20 & 60 & 1\myarcmin6 & 1.0 & FDF &  \\
   \#012 & SP & 4.64 & 8.53 & 5 & 15 & 60 & 5\myarcmin6 & 0.5 & CDFS &  \\
   \#013 & P  &      & 4.09 & 5 & 20 & 60 & & & CDFS &  \\
   \#014 & P  &      & 7.39 & 4 & 15 & 60 & & & CDFS &  \\
   \#015 & P  &      & 4.08 & 4 & 30 & 10 & & & CDFS &  \\
   \#016 & SP & 4.12 & 5.34 & 5 & 15 & 60 & 2\myarcmin4 & 2.0 & CDFS &  \\
   \#017 & S  & 4.02 &      & 2 & 30 & 50 & 2\myarcmin4 & 0.2 & CDFS & 0.135 \\
   \#018 & P  &      & 4.64 & 2 & 35 & 60 & & & CDFS & 0.146 \\
   \#019 & P  &      & 6.79 & 5 & 10 & 70 & & & CDFS &  \\
   \#020 & P  &      & 5.98 & 6 &    &    & & & CDFS &  \\
   \hline 
   \#021 & P  &      & 4.07 & 5 & 10 & 70 & & & DEEP2E &  \\
   \#022 & P  &      & 4.35 & 5 & 10 & 50 & & & DEEP2F &  \\
   \#023 & S  & 5.02 &      & 6 &    &    & 4\myarcmin8 & 0.2 & AM1 &  \\
   \#024 & S  & 4.05 &      & 4 & 25 & 90 & 9\myarcmin9 & 2.0 & AM1 &  \\
   \#025 & S  & 4.54 &      & 4 & 20 & 50 & 3\myarcmin2 & 0.2 & AM1 &  \\
   \#026 & S  & 4.10 &      & 5 & 10 & 50 & 2\myarcmin4 & 2.0 & Comp &  \\
   \#027 & P  &      & 4.35 & 5 & 10 & 60 & & & Comp &  \\
   \#028 & SP & 4.39 & 5.05 & 6 &    &    & 4\myarcmin0 & 0.1 & SHARC2 &  \\
   \#029 & S  & 4.12 &      & 2 & 30 & 90 & 2\myarcmin0 & 2.0 & SHARC2 &  \\
   \#030 & P  &      & 6.58 & 3 & 30 & 30 & & & A901 &  \\
   \hline 
   \#031 & P  &      & 4.51 & 5 & 10 & 40 & & & A901 &  \\
   \#032 & P  &      & 4.37 & 5 & 10 & 30 & & & A901 &  \\
   \#033 & P  &      & 7.06 & 2 & 35 & 30 & & & A901 &  \\
   \#034 & P  &      & 5.59 & 3 & 20 & 30 & & & A901 & 0.169* \\
   \#035 & S  & 4.63 &      & 3 & 15 & 50 & 6\myarcmin3 & 2.0 & A901 &  \\
   \#036 & P  &      & 7.08 & 2 & 45 & 60 & & & A901 & 0.161 \\
   \#037 & P  &      & 5.01 & 4 & 15 & 60 & & & A901 & \\
   \#038 & P  &      & 4.60 & 4 & 25 & 50 & & & A901 & \\
   \#039 & SP & 6.30 & 10.01 & 1 & 50 & 70 & 19\myarcmin8 & 0.1 & A901 & 0.161 \\
   \#040 & P  &      & 6.08 & 3 & 35 & 50 & & & A901 & 0.165 \\
   \hline 
   \#041 & P  &      & 4.52 & 4 & 20 & 80  & & & Pal3 & \\
   \#042 & S  & 4.13 &      & 3 & 25 & 40  & 2\myarcmin8 & 2.0 & Pal3 & 0.189* \\
   \#043 & P  &      & 5.14 & 4 & 20 & 40  & & & Pal3 & 0.094, 0.123, 0.215 \\
   \#044 & P  &      & 4.26 & 3 & 30 & 40  & & & CL1037-1243 &  \\
   \#045 & P  &      & 5.00 & 6 &    &     & & & CL1040-1155 &  \\
   \#046 & S  & 4.19 &      & 2 & 40 & 70  & 2\myarcmin4 & 0.5 & CL1040-1155 &  \\
   \#047 & P  & 3.53 & 4.06 & 5 & 5  & 120 & & & CL1040-1155 &  \\
   \#048 & S  & 4.07 &      & 6 &    &     & 9\myarcmin9 & 1.0 & CL1040-1155 &  \\
   \#049 & S  & 4.18 &      & 5 & 5  & 80  & 5\myarcmin6 & 0.1 & CL1054-1245 & 0.122* \\
   \#050 & S  & 4.07 &      & 6 &    &     & 3\myarcmin6 & 0.05 & CL1054-1146 &  \\
   \hline 
   \end{tabular}
   \normalsize
\end{table*}

\begin{table*}[t]
   \caption{\label{shearselclusters2}Shear-selected mass concentrations (part 2).}
   \small
   \begin{tabular}{lcrcrrrrrll}
   \noalign{\smallskip}
   \noalign{\smallskip}
   \hline 
   \hline 
   \noalign{\smallskip}
   \noalign{\smallskip}
   Peak & stat & \nus & \nup & class & rich & dist & scale & $x_{\mathrm{c}}$ & Field & z \\ 
   \noalign{\smallskip}
   \noalign{\smallskip}
   \hline 
   \noalign{\smallskip}
   \#051 & S  & 4.21 &      & 5 & 15 & 20  & 5\myarcmin6 & 0.1 & CL1059-1253 &  \\
   \#052 & P  &      & 4.60 & 5 & 20 & 30  & & & CL1059-1253 &  \\
   \#053 & S  & 4.02 &      & 4 & 20 & 120 & 6\myarcmin3 & 2.0 & CL1059-1253 & 0.41, 0.52, 0.60 \\
   \#054 & S  & 4.01 &      & 4 & 15 & 30  & 2\myarcmin4 & 2.0 & CL1059-1253 &  \\
   \#055 & SP & 4.53 & 6.31 & 5 & 5  & 20  & 3\myarcmin2 & 2.0 & DEEP3D &  \\
   \#056 & P  &      & 4.08 & 4 & 15 & 10  & & & DEEP3D & \\
   \#057 & SP & 4.24 & 7.48 & 4 & 25 & 80  & 5\myarcmin6 & 0.1 & DEEP3D &  \\
   \#058 & SP & 4.38 & 5.04 & 4 & 20 & 70  & 5\myarcmin6 & 0.5 & CL1119-1129 &  \\
   \#059 & S  & 4.25 &      & 6 &    &     & 1\myarcmin6 & 0.1 & DEEP3D &  \\
   \#060 & P  &      & 5.82 & 6 &    &     & & & DEEP3D & \\
   \hline 
   \#061 & S  & 4.08 &      & 4 & 20 & 40  & 2\myarcmin0 & 2.0 & DEEP3A &  \\
   \#062 & S  & 4.21 &      & 4 & 20 & 70  & 3\myarcmin2 & 0.2 & DEEP3A &  \\
   \#063 & P  &      & 4.67 & 5 & 20 & 120 & & & CL1138-1133 & \\
   \#064 & P  &      & 4.31 & 4 & 15 & 60  & & & CL1138-1133 & \\
   \#065 & P  &      & 4.97 & 4 & 30 & 50  & & & CL1138-1133 & \\
   \#066 & SP & 4.02 & 5.48 & 3 & 25 & 50  & 7\myarcmin9 & 0.05 & A1347\_P1 &  \\
   \#067 & P  &      & 4.10 & 5 & 35 & 50  & & & A1347\_P3 & \\
   \#068 & S  & 4.00 &      & 5 & 5  & 50  & 1\myarcmin6 & 0.1 & A1347\_P1 &  \\
   \#069 & SP & 4.21 & 4.95 & 5 & 10 & 60  & 6\myarcmin3 & 0.025 & A1347\_P3 &  \\
   \#070 & SP & 4.32 & 4.94 & 6 &    &     & 7\myarcmin1 & 0.05 & A1347\_P3 &  \\
   \hline 
   \#071 & S  & 4.28 &      & 4 & 20 & 50 & 2\myarcmin4 & 2.0 & S11 &  \\
   \#072 & P  &      & 4.37 & 4 & 15 & 90 & & & A1347\_P1 & \\
   \#073 & P  &      & 6.17 & 3 & 25 & 70 & & & A1347\_P1 & \\
   \#074 & S  & 4.27 &      & 4 & 40 & 60 & 1\myarcmin6 & 0.5 & S11 &  \\
   \#075 & S  & 4.43 &      & 3 & 35 & 30 & 1\myarcmin6 & 0.5 & S11 &  \\
   \#076 & S  & 4.27 &      & 6 &    &    & 2\myarcmin8 & 2.0 & S11 &  \\
   \#077 & S  & 4.76 &      & 4 & 20 & 70 & 13\myarcmin9 & 0.05 & S11 &  \\
   \#078 & SP & 4.00 & 4.05 & 2 & 40 & 40 & 2\myarcmin8 & 2.0 & S11 & 0.119 \\
   \#079 & P  &      & 4.02 & 3 & 30 & 40 & & & A1347\_P4 & \\
   \#080 & S  & 4.02 &      & 4 & 20 & 10 & 2\myarcmin8 & 0.1 & A1347\_P4 &  \\
   \hline 
   \#081 & P  &      & 4.78 & 6 &    &     & & & A1347\_P2 & \\
   \#082 & SP & 4.87 & 5.83 & 1 & 60 & 80  & 13\myarcmin9 & 0.025& S11 & 0.106 \\
   \#083 & P  &      & 5.33 & 5 & 15 & 60  & & & CL1202-1224 & \\
   \#084 & P  &      & 4.61 & 5 & 40 & 120 & & & CL1202-1224 & 0.4 \\
   \#085 & S  & 4.63 &      & 5 & 10 & 50  & 2\myarcmin4 & 0.5 & CL1216-1201 &  \\
   \#086 & S  & 4.43 &      & 5 & 10 & 90  & 4\myarcmin0 & 0.5 & CL1216-1201 &  \\
   \#087 & S  & 4.01 &      & 5 & 30 & 130 & 5\myarcmin6 & 2.0 & CL1216-1201 & 0.33 \\
   \#088 & S  & 4.64 &      & 6 &    &     & 3\myarcmin2 & 0.2 & CL1216-1201 &  \\
   \#089 & P  &      & 5.01 & 5 & 10 & 70  & & & CL1301-1139 & \\
   \#090 & S  & 4.18 &      & 3 & 30 & 70  & 9\myarcmin9 & 2.0 & CL1301-1139 &  \\
   \hline 
   \#091 & SP & 4.64 & 4.99 & 3 & 30 & 50 & 11\myarcmin9 & 0.025& CL1353-1137 &  \\
   \#092 & P  &      & 5.28 & 3 & 35 & 50 & & & CL1353-1137 & \\
   \#093 & SP & 4.39 & 4.40 & 5 & 15 & 90 & 3\myarcmin6 & 2.0 & C0400 &  \\
   \#094 & S  & 4.14 &      & 6 &    &    & 5\myarcmin6 & 2.0 & C0400 &  \\
   \#095 & P  &      & 5.30 & 6 &    &    & & & C04p2 & \\
   \#096 & P  &      & 4.32 & 6 &    &    & & & C04p1 & \\
   \#097 & P  &      & 4.15 & 6 &    &    & & & C04m1 & \\
   \#098 & SP & 4.64 & 7.60 & 4 & 25 & 20 & 5\myarcmin6 & 2.0 & C04p2 &  \\
   \#099 & S  & 4.22 &      & 6 &    &    & 2\myarcmin4 & 2.0 & C04m4 &  \\
   \#100 & P  & 5.13 & 5.90 & 6 &    &    & & & C04p3 & \\
   \hline 
   \end{tabular}
   \normalsize
\end{table*}

\begin{table*}[t]
   \caption{\label{shearselclusters3}Shear-selected mass concentrations (part 3).}
   \small
   \begin{tabular}{lcrcrrrrrll}
   \noalign{\smallskip}
   \noalign{\smallskip}
   \hline 
   \hline 
   \noalign{\smallskip}
   \noalign{\smallskip}
   Peak & stat & \nus & \nup & class & rich & dist & scale & $x_{\mathrm{c}}$ & Field & z \\ 
   \noalign{\smallskip}
   \noalign{\smallskip}
   \hline 
   \noalign{\smallskip}
   \#101 & P  &      & 5.59 & 6 &    &    & & & CL1420-1236 &  \\
   \#102 & P  &      & 5.59 & 6 &    &    & & & C04m4 & \\
   \#103 & P  &      & 4.55 & 6 &    &    & & & C04m4 & \\
   \#104 & P  &      & 4.90 & 4 & 15 & 80 & & & C04m4 & \\
   \#105 & S  & 4.08 &      & 6 &    &    & 2\myarcmin4 & 2.0 & CL1420-1236 &  \\
   \#106 & S  & 4.07 &      & 6 &    &    & 6\myarcmin3 & 2.0 & CL1420-1236 &  \\
   \#107 & P  &      & 4.12 & 6 &    &    & & & CL1420-1236 &  \\
   \#108 & S  & 4.43 &      & 6 &    &    & 2\myarcmin4 & 1.0 & C04m3 &  \\
   \#109 & SP & 4.03 & 6.29 & 5 & 10 & 20 & 6\myarcmin3 & 0.2 & CL1420-1236 &  \\
   \#110 & P  &      & 4.02 & 3 & 35 & 70 & & & C04p3 & \\
   \hline 
   \#111 & S  & 4.05 &      & 5 & 30 & 90 & 4\myarcmin0 & 0.1 & CL1420-1236 &  \\
   \#112 & S  & 4.01 &      & 6 &    &    & 4\myarcmin0 & 0.05 & C04m3 &  \\
   \#113 & S  & 4.07 &      & 6 &    &    & 2\myarcmin4 & 0.5 & FIELD17\_P1 &  \\
   \#114 & P  &      & 4.59 & 2 & 80 & 60 & & & FIELD17\_P3 & \\
   \#115 & P  &      & 4.59 & 3 & 25 & 50 & & & FIELD17\_P3 & \\
   \#116 & S  & 4.28 &      & 6 &    &    & 1\myarcmin6 & 2.0 & FIELD17\_P1 &  \\
   \#117 & P  &      & 4.60 & 5 & 10 & 30 & & & FIELD17\_P3 & \\
   \#118 & P  &      & 4.97 & 5 & 10 & 90 & & & FIELD17\_P3 & \\
   \#119 & P  &      & 4.35 & 6 &    &    & & & FIELD17\_P3 & \\
   \#120 & P  &      & 5.76 & 3 & 25 & 40 & & & FIELD17\_P3 & 0.132* \\
   \hline 
   \#121 & S  & 4.11 &      & 5 & 20 & 70 & 2\myarcmin0 & 0.5 & FIELD4\_P1 &  \\
   \#122 & S  & 4.09 &      & 4 & 20 & 40 & 3\myarcmin6 & 0.2 & FIELD4\_P1 &  \\
   \#123 & S  & 4.02 &      & 3 & 35 & 40 & 2\myarcmin0 & 0.2 & FIELD4\_P3 &  \\
   \#124 & S  & 4.00 &      & 4 & 15 & 10 & 2\myarcmin4 & 1.0 & FIELD4\_P3 &  \\
   \#125 & P  &      & 4.35 & 6 &    &    & & & FIELD4\_P4 & \\
   \#126 & S  & 4.02 &      & 5 & 5  & 30 & 2\myarcmin0 & 2.0 & FIELD4\_P3 &  \\
   \#127 & P  &      & 4.46 & 6 &    &    & & & FIELD4\_P4 & \\
   \#128 & SP & 4.13 & 4.21 & 3 & 30 & 60 & 7\myarcmin9 & 0.025 & FIELD4\_P4 &  \\
   \#129 & S  & 4.00 &      & 5 & 15 & 70 & 3\myarcmin2 & 0.1 & FIELD4\_P4 &  \\
   \#130 & SP & 5.10 & 7.61 & 4 & 15 & 80 & 17\myarcmin8 & 0.05 & B8p0 &  \\
   \hline 
   \#131 & SP & 4.29 & 4.30 & 5 & 10 & 90 & 19\myarcmin8 & 0.05 & B8m1 &  \\
   \#132 & S  & 4.12 &      & 5 & 10 & 70 & 2\myarcmin4 & 1.0 & B8m2 &  \\
   \#133 & P  &      & 4.05 & 6 &    &    & & & B8m3 & \\
   \#134 & SP & 4.05 & 6.06 & 6 &    &    & 7\myarcmin1 & 2.0 & B8p0 &  \\
   \#135 & P  &      & 4.14 & 6 &    &    & & & B8m2 & \\
   \#136 & SP & 4.45 & 10.20 & 3 & 30 & 50 & 2\myarcmin8 & 0.5 & B8p0 &  \\
   \#137 & SP & 4.12 & 6.26 & 6 &    &    & 3\myarcmin6 & 2.0 & B8p0 &  \\
   \#138 & P  &      & 4.70 & 5 & 15 & 60 & & & B8p0 &  \\
   \#139 & SP & 4.35 & 6.95 & 4 & 15 & 10 & 2\myarcmin0 & 0.2 & B8p0 &  \\
   \#140 & S  & 4.13 &      & 4 & 15 & 90 & 15\myarcmin9 & 0.05 & B8p2 & 0.07* \\
   \hline 
   \#141 & SP & 4.59 & 7.76 & 2 & 40 & 80  & 3\myarcmin2 & 0.5 & B8p0 &  \\
   \#142 & S  & 4.71 &      & 3 & 35 & 70  & 4\myarcmin0 & 2.0 & B8p1 & 0.318 \\
   \#143 & S  & 4.07 &      & 6 &    &     & 6\myarcmin3 & 0.1 & B8p3 &  \\
   \#144 & P  &      & 4.71 & 2 & 40 & 80  & & & B8p0 & \\
   \#145 & S  & 4.25 &      & 6 &    &     & 1\myarcmin6 & 0.1 & B8m1 &  \\
   \#146 & S  & 4.11 &      & 6 &    &     & 3\myarcmin6 & 2.0 & B8p1 &  \\
   \#147 & P  &      & 6.78 & 2 & 90 & 100 & & & B8p0 & \\
   \#148 & P  &      & 4.31 & 4 & 15 & 10  & & & B8p0 & \\
   \#149 & S  & 4.27 &      & 5 & 10 & 0   & 2\myarcmin0 & 0.5 & B8m1 &  \\
   \#150 & P  &      & 4.21 & 4 & 30 & 80  & & & B8p0 & \\
   \hline 
   \#151 & P  &      & 4.01 & 6 &    &     & & & B8m2 & \\
   \#152 & SP & 4.14 & 4.20 & 5 & 10 & 30  & 2\myarcmin0 & 1.0 & B8p0 &  \\
   \#153 & P  &      & 4.12 & 5 & 25 & 100 & & & DEEP1C & \\
   \#154 & P  &      & 4.33 & 6 &    &     & & & DEEP1E & \\
   \#155 & S  & 4.02 &      & 5 & 20 & 120 & 3\myarcmin6 & 0.5 & DEEP1C & 0.135* \\
   \#156 & P  &      & 6.22 & 6 &    &     & & & DEEP1A & \\
   \#157 & S  & 4.35 &      & 4 & 20 & 30  & 7\myarcmin9 & 0.05 & DEEP1A & 0.151, 0.176 \\
   \#158 & P  &      & 6.91 & 2 & 40 & 30  & & & DEEP1A & 0.175 \\
   \hline 
   \end{tabular}
   \normalsize
\end{table*}

\begin{table*}
   \caption{\label{surveydiffs}Bright (classes 1 to 4) and dark (5-6) peaks
   for the various survey data sources. The columns contain the data source,
   the number of bright and dark peaks, the ratio between dark and bright
   peaks, the average exposure times and image seeing, the area covered, plus
   the number of bright and dark peaks per unit area. For the COMBO-17 field we
   give in parenthesis the corresponding values when the known structures are
   subtracted.} 
   \begin{tabular}{lcccccccc}
   \noalign{\smallskip}
   \noalign{\smallskip}
   \hline 
   \hline 
   \noalign{\smallskip}
   \noalign{\smallskip}
   Source & $n_{\rm b}$ & $n_{\rm d}$ & $n_{\rm d}/n_{\rm b}$ 
   & t[ksec] & s[ \myarcsecnodot] & A [$\,\Box^2$] & $n_{\rm b}/A$ & 
   $n_{\rm d}/A$ \\ 
   \noalign{\smallskip}
   \noalign{\smallskip}
   \hline 
   \noalign{\smallskip}
   EDisCS      & 10  & 21 & 2.1 & 3.5  & 0.90 & 3.8 & 2.6 & 5.5 \\
   ASTROVIRTEL & 18  & 27 & 1.5 & 4.9  & 0.87 & 6.1 & 3.0 & 4.4 \\
   COMBO-17    & 25 (18) & 13 & 0.5 (0.7) & 25.7 & 0.80 & 1.6 & 15.6 (11.2) & 8.1 \\
   EIS         & 6   & 7  & 1.2 & 6.8  & 0.94 & 2.9 & 2.1 & 2.4 \\
   Own obs.    & 13  & 18 & 1.4 & 8.9  & 0.88 & 4.2 & 3.1 & 4.3 \\
   \hline 
   \end{tabular}
   \normalsize
\end{table*}

\begin{table*}
   \caption{\label{fractionbrightdark}Number of bright (classes 1-4) and 
     dark (5-6) peaks and their ratios for the $S$- and the
     $P$-statistics.}
   \begin{tabular}{lccc|ccc}
   \noalign{\smallskip}
   \noalign{\smallskip}
   \hline 
   \hline 
   \noalign{\smallskip}
   \noalign{\smallskip}
   $S/N$ & $n_{\rm b}(S)$ & $n_{\rm d}(S)$ & $n_{\rm d}/n_{\rm b}$  & 
           $n_{\rm b}(P)$ & $n_{\rm d}(P)$ & $n_{\rm d}/n_{\rm b}$ \\ 
   \noalign{\smallskip}
   \noalign{\smallskip}
   \hline 
   \noalign{\smallskip}
   $4.0-4.25$ & 23 & 31 & 1.3 & 7  & 12 & 1.7 \\
   $4.25-4.5$ & 9  & 11 & 1.2 & 5  & 9  & 1.8 \\
   $4.5-4.75$ & 6  &  5 & 0.8 & 6  & 7  & 1.2 \\
   $4.75-5.0$ & 2  &  0 & 0.0 & 3  & 4  & 1.3 \\
   $5.0-6.0$  & 2  &  1 & 0.5 & 8  & 11 & 1.4 \\
   $>6.0$     & 1  &  0 & 0.0 & 15 & 7  & 0.5 \\
   \hline 
   total      & 43 & 48 & & 44 & 50 & \\
   \hline 
   \end{tabular}
   \normalsize
\end{table*}

\begin{figure*}[t]
\includegraphics[width=1.0\hsize]{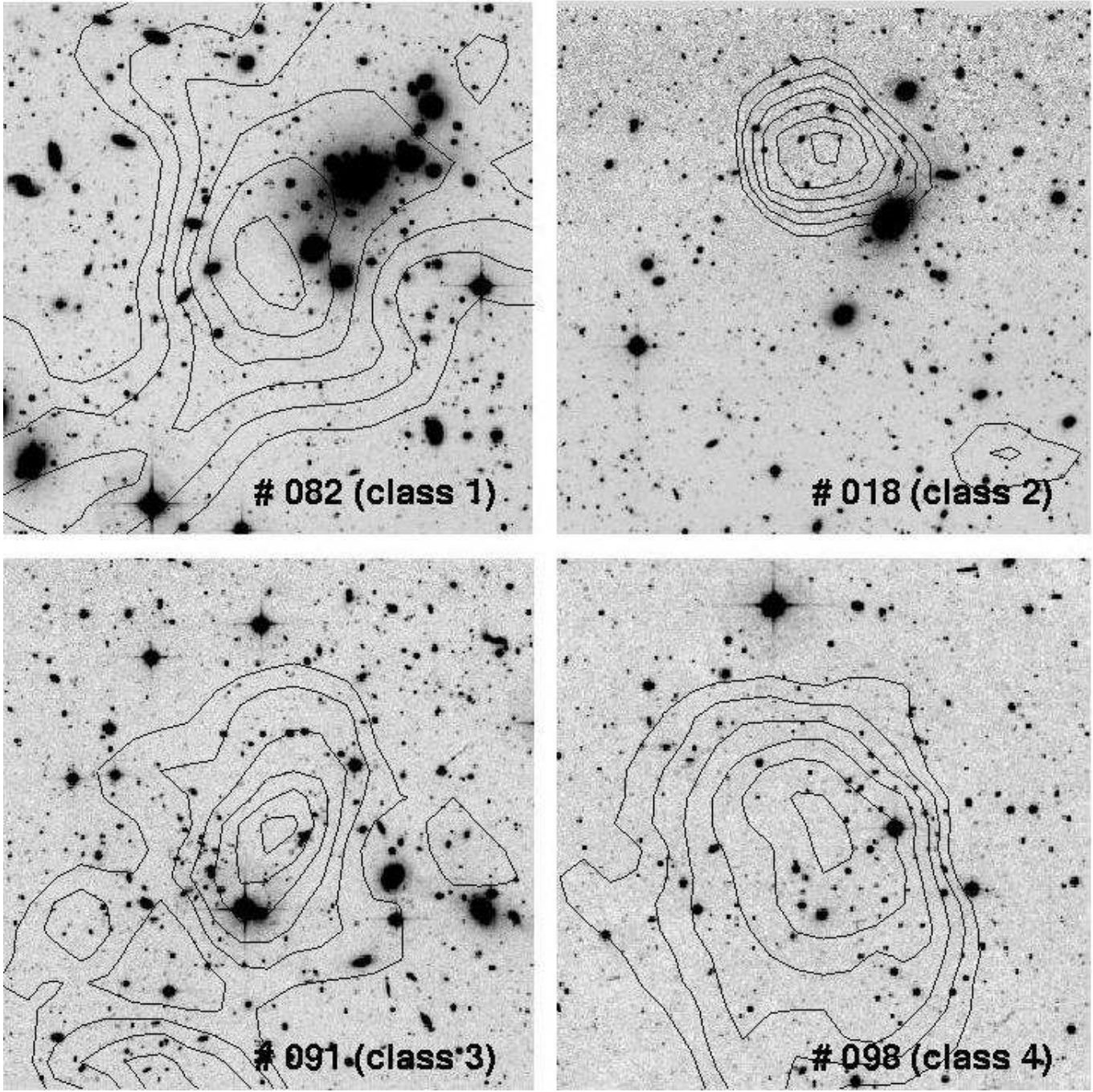}
\caption{\label{classes}{Illustrates the typical appearance of bright clusters
  of various classes, as defined in Sect. \ref{defclass}. Note that the
  resolution is in general not high enough to distinguish smaller member
  galaxies from stars. The field of view is 4\myarcmin3.}}
\end{figure*}

\end{document}